\documentclass{amsart}
\usepackage{amssymb}
\usepackage{latexsym}
\title[Closed/open conformal field theories]{Closed and open
conformal field theories and their anomalies}
\author{Po Hu and Igor Kriz}
\input xypic
\newtheorem{theorem}{Theorem}

\newtheorem{lemma}[theorem]{Lemma}

\def\Proof{\medskip\noindent{\bf Proof: }}

\def\Z{\mathbb{Z}}

\def\C{\mathbb{C}}

\def\R{\mathbb{R}}
\def\C{\mathbb{C}}

\def\N{\mathbb{N}}

\def\Pi{\mathbb{P}^{\infty}}

\def\umh{\underline{\mh}}
\def\ct{\mathbb{C}^{\times}}

\def\qed{\hfill$\square$\medskip}

\def\Zpk{\mathbb{Z}/p^{k}}
\def\Zpk1{\mathbb{Z}/p^{k-1}}

\newcommand{\rref}[1]{(\ref{#1})}

\newcommand{\cform}[3]{\begin{array}{c}
{\scriptstyle #3}\\
#1\\
{\scriptstyle #2}\end{array}}

\newcommand{\beg}[2]{\begin{equation}\label{#1}#2\end{equation}}
\def\r{\rightarrow}
\def\mc{\mathcal{C}}
\def\mb{\mathcal{B}}
\def\mf{\mathcal{F}}
\def\mh{\mathcal{H}}

\def\sl2{\widetilde{SL_{2}(\Z)}}

\begin{document}

\maketitle

\vspace{5mm}

\section{Introduction}

The main purpose of this paper is to give rigorous mathematical
foundations for investigating closed and closed/open conformal
field theories (CFT's) and their anomalies. In physics, the topic
of closed/open CFT
has been extensively discussed in the literature (see e.g.
\cite{doug,laz2,laz3,diac}).
Our investigation was originally inspired by two sources: Edward Witten
(cf. \cite{w}) proposed a general program
for using $K$-theory to classify stable $D$-branes in string
theory. On the other hand, Moore and Segal \cite{ms} obtained a mathematically
rigorous approach to classifying $D$-branes in the case of 
$2$-dimensional topological
quantum field theory (TQFT).

\vspace{3mm}
We attempted to consider a case in between, namely $D$-branes in conformal
field theory. There are many reasons for $D$-branes to be easier in CFT
than in string theory. First of all, one does not have to insist on anomaly
cancellation, and can investigate the anomaly instead. Another reason
is that CFT makes good sense even without supersymmetry, while string theory
does not. (In fact, in this paper, we restrict attention to non-supersymmetric
CFT, although mostly just
in the interest of simplicity.) Most importantly, however, any CFT 
approach to string theory amounts to looking at the complicated string
moduli space only through the eyes of one tangent space, which is
a substantial simplification. We will make
some comments on this relationship between CFT and string theory
in Section \ref{s4a} below.

\vspace{3mm}
On the other hand, CFT is incomparably more complicated than $2$-dimensional
TQFT. Because of this, in fact, a theorem classifying $D$-branes (as outlined in
\cite{ms} for TQFT) seems, at the present time, out of reach
for CFT. However, an exact mathematical {\em definition} of the
entire structure of closed/open CFT is a reasonable goal which we do
undertake here. To this end, we use our formalism of stacks of lax commutative
monoids with cancellation (SLCMC's), developed in \cite{hk}, and
reviewed in the Appendix. Also,
we describe analogues of some of the concepts of \cite{ms} for CFT's, and
observe some interesting new phenomena. For example, the 
correct generalization of module over the algebra corresponding to TQFT,
is not a VOA module, but a different kind of module (which we
call {\em $D$-brane module}). When specializing to genus $0$ surfaces, 
this leads to a generalization of the notion of module over an operad.
The $K$-theory of such modules (with some finiteness assumption)
is therefore a candidate for a CFT analogue of the $K$-theory considered
in \cite{ms} (CFT ``$D$-brane cohomology''). These topics are discussed in
Section \ref{s2} below. 

\vspace{3mm}
In Section \ref{s4}, we give a basic example, the free bosonic CFT
(=linear $\sigma$-model), and show how to obtain the $D$-brane modules
corresponding to Von Neumann and Dirichlet boundary conditions for open
strings. As we will see, however, even in this basic case,
a substantial complication
is giving a mathematically rigorous treatment of the convergence issues
of the CFT.

\vspace{3mm}
Up to this point, we suppressed the discussion of anomaly,
by assuming that anomaly is $1$-dimensional. 
However, there is an obvious suggestion a parallel between 
the set of $D$-branes
of a closed/open CFT, and the set of labels of a modular functor
of a rational CFT (RCFT, see \cite{ms1,ms2,ms3,scft}). It therefore seems
we should look for axioms for the most general possible kind of anomaly
for closed/open CFT, which would
include sets of both $D$-branes and modular functor labels.

\vspace{3mm}
There are, however, further clues which suggests that the notion of ``sets''
in this context is too restrictive. Notably, the free $\C$-vector
space $\C S$ on the set of labels $S$
of a modular functor is the well known Verlinde algebra \cite{verlinde}.
But the multiplication rule of
the Verlinde algebra uses only of {\em dimensions} of vector
spaces involved in the modular functor, so it seems that if one wants
to consider the spaces themselves, it is that one should consider,
instead of $\C S$, the {\em free $2$-vector space on $S$}). Is it
possible to axiomatize modular functors for RCFT's in a way which
uses $2$-vector spaces in place of sets of labels? 

\vspace{3mm}
In Section \ref{s5}, we answer this last question in the affirmative.
This is rather interesting, because it leads to other questions:
the authors \cite{hk} previously proposed RCFT as a possible
tool for geometrically modelling elliptic cohomology, while Baas-Dundas-Rognes
\cite{rog} obtained a version of elliptic cohomology based on $2$-vector
spaces. Is there a connection? Observations made in \cite{rog} show that
the right environment of such discussion would be a suitable {\em group
completion} of the symmetric bimonoidal category $\C_{2}$ of vector spaces, 
while simultaneously noting the necessary difficulty of any such construction.
Nevertheless, we propose in, Section \ref{s6a}, such group completion,
despite major technical difficulties.
Our construction involves super-vector spaces, and thus suggests
further connections with P. Deligne's observation \cite{spin} that the
modulars functor of $bc$-systems must be considered as super-vector spaces,
and with the work of Stolz-Teichner \cite{st}, who, in their approach to
elliptic objects, also noticed the role of fermions and, in effect, 
what amounts
$1$-dimensional super-modular functors.

\vspace{3mm}
However, let us return to $D$-branes. Is it possible to formulate axioms
for anomalies of closed/open CFT's analogous to the $2$-vector space 
approach to modular functors? We give, again, an
affirmative answer, although another surprise awaits us here: while
the ``set of labels'' of a modular functor was naturally a $2$-vector
space, the ``set of $D$-branes'' of a closed/open CFT must be a $3$-vector
space! We discuss this, and
propose axioms for a general anomaly of closed/open CFT
in Section \ref{s7}. An intriguing question is whether the group completion approach
of Section \ref{s6a} can also be extended to the case closed/open CFT anomaly.

\vspace{3mm}
\noindent
{\bf Acknowledgement:} The authors would like to think N.A.Baas and J.Rognes
for their helpful comments, and for detecting mistakes in the original version
of this paper.

\vspace{5mm}

\section{Closed/open CFT's with $1$-dimensional anomaly, $D$-brane modules
and $D$-brane cohomology}
\label{s2}

There is substantial physical literature on the subject of $D$-branes
(see e.g. \cite{doug,laz2,laz3,diac}).
In this paper, we shall discuss a mathematically rigorous approach to
$D$-branes in non-supersymmetric CFT's. Moreover, in Sections
\ref{s2}-\ref{s4},
we shall restrict attention to $1$-dimensional
anomaly allowed both on the closed CFT and the $D$-brane.
More advanced settings will be left to the later sections.

\vspace{3mm}
We begin by defining the stack of lax commutative monoids with
cancellation (SLCMC) corresponding to oriented open/closed
string (more precisely 
conformal field) theory. SLCMC's were introduced in \cite{hk},
but to make this paper self-contained, we review
all the relevant definitions
in the Appendix. We consider a set $L$.
This is not our set of labels, it is the set of $D$-branes.
Our set of labels consists of
$K^{\prime}=L\times L$ 
which we will call {\em open labels} and we will put $K=K^{\prime}
\coprod \{1\}$ where $1$ is the {\em closed label}. 

\vspace{3mm}
To define the SLCMC $\mathcal{B}$
for oriented open/closed CFT, we shall first define the LCMC of its
sections over a point. As usual (see \cite{hk}), the underlying lax commutative
monoid is the category of finite sets labelled by a certain set
$K$, not necessarily finite.
Before describing the exactly correct analytic and conformal structure,
we first specify that
these are compact oriented
surfaces ($2$-manifolds) $X$ together with homeomorphic
embeddings $c_i:S^{1}\r \partial X$, $d_j:I\r\partial X$ with disjoint
images. Moreover, each $c_i$ is labelled with $1$, and each $d_j$ is labelled
by one of the open labels $K^{\prime}$. Moreover,
each connected component of 
$$\partial X - \bigcup Im(c_i)-\bigcup Im(d_j)$$
(which we shall call $D$-brane components)
is labelled with an element of $L$, and each $d_j$ is labelled with
the pair $(\ell_{1}, \ell_{2})\in L\times L$
of $D$-branes which the beginning point and endpoint of $d_j$ abut.
The $c_i$'s and $d_j$'s are considered
inbound or outbound depending on the usual comparison of their orientation
with the orientation of $X$. 

\vspace{3mm}
It is now time to describe the smoothness and conformal structure on $X$.
To this end, we simply say that $X$ is a smooth complex $1$-manifold with analytic
(real) boundary and corners; this means that, the interior of $X$ is a complex
$1$-manifold, and the neighbourhood of a boundary point $x$ of $X$ is modeled
by a chart which whose source is either an open subset of the halfplane $\mathbb{H}=
\{z\in\C| Im(z)\geq 0\}$ where $0$ maps to $x$
or an open subset of
the quadrant $\mathbb{K}=\{z\in\C| Im(z)\geq 0, Re(z)\geq 0\}$ where
$0$ maps to $x$; the
transition maps are (locally) holomorphic maps which extend biholomorphically
to an open
neighbourhood of $0$ in $\C$. The points of the boundary whose neighbourhoods
are modelled by open neighborhoods of $0$ in $\mathbb{K}$ are called
{\em corners}.
Further, we specify exactly which
points are the corners of an open/closed string world sheet: 
we require that the corners be precisely the endpoints of the images of
open string boundary parametrizations. We also require that open as
well as closed string parametrizations be real-analytic diffeomorphisms
onto their image; this completes the definition of the objects.

\vspace{3mm}
An isomorphism $X\r Y$ is a diffeomorphism which preserves complex structure,
$D$-brane labels, 
is smooth on the interior, and commutes with the $c_i$, $d_j$ (which
we shall call parametrization components - note that the set of
parametrization
components is not ordered, so an automorphism may switch them).

\vspace{3mm}
Now to define the SLCMC $\mathcal{B}$, the main issue is fixing
the Grothendieck topology. We use simply finite-dimensional smooth manifolds with
open covers. As usual, the underlying stack of lax commutative monoids
is the stack of covering spaces with locally constant $K$-labels
(analogously to \cite{hk}). Sections of $\mathcal{B}$ over $M$
are smooth manifolds fibered over $M$, where the fibers are elements of
$\mathcal{B}$, and the structure varies smoothly in the obvious sense.
It is important, however, to note that it does not seem possible
to define this stack over the Grothendieck topology of {\em complex}
manifolds and open covers; in other words, it does not appear
possible to discuss {\em chiral} CFT's with $D$-branes. To see this,
we consider the following 

\vspace{3mm}
\noindent
{\bf Example:} The moduli space of elliptic curves $E$ with an unparametrized
hole (i.e. one closed $D$-brane component with a given
label). It is easily seen that the
moduli space of such worldsheets is the ray $(0,\infty)$, i.e. not a complex
manifold. To see this, the key point is to notice that the invariant
$Im(\tau)$ of the elliptic curve $F$ obtained by attaching a unit disk to $E$
along the $D$-brane component does not depend on its parametrization.
Intuitively, this seems plausible since $Im(\tau)$ is the ``volume''.
To rigorize the argument, we first recall that if one cuts the elliptic
curve along a non-separating curve, then $Im(\tau)$ can be characterized
as the ``thickness'' of the resulting annulus (every annulus is 
conformally equivalent to a unique annulus of the form $S^{1}\times [0,r]$
for some boundary parametrization; $r$ is the thickness). But now any
reparametrization of the $D$-brane component $c$ of $E$ is a composition of
reparametrizations which are identity outside of a certain small interval
$J\subset c$. Thus, it suffices to show that the invariant $Im(\tau)$ does
not change under such reparametrizations. However, we can find a smooth
non-separating curve $d\supset J$ in $F$; then cutting $F$ along $d$,
the said change of parametrization of $c$ becomes simply a change of
parametrization of one of the boundary components of $F$; we already
know that does not affect thickness.

\vspace{3mm}
By a {\em $K$-labelled closed/open CFT} ($D$-brane category) with $1$-dimensional
anomaly $(H_i)_{i\in K}$ we shall mean a CFT with $1$-dimensional
modular functor on the SLCMC $\mb$ over the stack of lax monoids $S_{K}$,
with target in the SLCMC $\underline{H}_{K}$. This means a lax morphism
of SLCMC's
$$\tilde{\mathcal{B}}\r (\underline{H}_i)_{i\in K}$$
where $\tilde{\mb}$ is a $\ct$-central extension of $\mb$.
These concepts were defined in \cite{hk} 
(see the Appendix for a review).

\vspace{3mm}
From this point of view, we see that the question
of classifying $D$-branes for a given (closed) CFT $H$ is, at least
a priori, not well stated:
it is possible that there may be different $D$-brane categories on the
same set of objects, or even with the same endomorphism monoids (the state
spaces $H_{aa}$), but with different choices of $H_{ab}$. One may therefore
ask for a classification of all possible $D$-brane
categories for a given closed CFT. 

We shall, however, not follow that route in this Section. 
Instead, we would like to capture
the more basic physical intuition that $D$-branes should be objects with
their own properties, which determine in some canonical way the morphisms
between them. Therefore, we shall restrict our attention, 
instead of classifying
all possible $D$-brane categories, to attempting to build one canonical $D$-brane
category.

\vspace{3mm}
For a hint of what such category would have to look like, we follow the
approach of Moore-Segal \cite{ms}. They suggested that, in the case of
$2$-dimensional topological quantum field theory (TQFT), $D$-branes
should be, at least under some assumptions, (finitely generated projective) 
modules over the algebra $A$ associated with the TQFT.
Therefore, they reach the conclusion that $D$-branes in TQFT are classified
by $K^{0}(A)$, although of course more precisely $K^{0}(A)$ is
the Grothendieck group of the category of $D$-branes and isomorphisms.

\vspace{3mm}
It is very appealing to try to find an analogue of the same treatment
for $D$-branes in conformal field theory. In particular, one can ask what
is the right notion of ``module'' over a CFT which would correspond to
$D$-branes. It is important to note that the right notion does
not seem to be the CFT analogue of modules over a VOA, which play
an important role in the investigation of chiral CFT's (since they
determine their modular functor). However, the $D$-brane classification
question can be asked even for CFT's with $1$-dimensional
anomaly, and, as we already saw, $D$-branes on the other
hand cannot be
chiral. While we don't think the analogues of VOA modules for non-chiral
theories have been investigated, the analogy with the chiral case
suggests that one would see only one module over any CFT with $1$-dimensional
anomaly, and that, in any case, this notion of module seems to be the
wrong one. 

\vspace{3mm}
We shall next propose what seems like a simplest possible 
notion of ``$D$-brane module''. Therefore, 
we also construct an ``$D$-brane cohomology'' which is the $K$-theory
of this category of $D$-brane modules. While we do not investigate
tensor products in this paper explicitly, it
should however be noted  
that our $D$-brane cohomology
is a {\em module} over ordinary $K$-theory in the
sense of algebraic topology; this means that
$D$-brane cohomology should be, in itself, considered a kind
of $K$-theory; it does not appear to contain the type of
homotopy-theoretical
information which we would expect, for example, in elliptic cohomology.
It also does not appear to have the type of modularity associated with elliptic
cohomology.

Our notion of $D$-brane module captures the simplest possible feature required
of open string theory - propagation along a single $D$-brane, with built in
correlation with closed CFT events. Mathematically, this is done as follows.

\vspace{3mm}
To define a $D$-brane module, we must introduce
a certain special partial SLCMC $\mathcal{P}$. Namely, we will consider two labels, $1$
(closed) and $m$ (open, or ``module''). We will consider open string worldsheets $X$.

However, we shall impose the following special condition: The $D$-brane components
of $X$ are unlabelled. In addition, in each connected
component of the worldsheet, there are no closed $D$-brane components,
and there are either no $D$-brane comoponents at all,
or alternately precisely two open string 
parametrization components $\alpha$ and $\omega$, 
which lie on the same boundary component, and
moreover $\alpha$ is oriented inbound and $\omega$ outbound.
We let these be the LCMC forming
the sections over a point of the partial SLCMC $\mathcal{P}$;
(it is extended to an SLCMC in the canonical way described in \cite{hk}.

Note that $\mathcal{P}$ is ``almost an SLCMC'': the only reason it is
partial is that we do not allow to glue the inbound and outbound open
string component of the same connected component together, which would
produce two closed $D$-brane components, which we do not allow.

\vspace{3mm}
Now a $D$-brane module $H_{a}$ over a closed CFT $H$ is defined to be
an abstract CFT with $1$-dimensional anomaly
over the partial SLCMC $\mathcal{P}$, where the Hilbert space assigned
to label $1$, resp. $m$ is $H$, resp. $H_{a}$. Note that it makes sense
to define a sub-$D$-brane module, and that, of course, a set-theoretical intersection
of an arbitrary set of $D$-brane modules is again an $D$-brane module.
We can, therefore, speak of an $D$-brane module $H_{a}$
spanned by a set of elements of $H_{a}$. We say that an $D$-brane module
is finitely generated if it is spanned by a finite set of elements.
Note, also, that isomorphism of $D$-brane modules is well defined
in the obvious way (due to the fact that the SLCMC's of Hilbert spaces
are ``strict'', in the sense that the only laxness comes from the
underlying lax-commutative monoid of sets). It is slightly less
obvious that $D$-brane modules with the same $1$-dimensional
anomaly also have a well defined direct sum:
to get the sum of $H_{a}$ and $H_{b}$,
the Hilbert space attached to the label $1$ remains $H$, but the
Hilbert space attached to $m$ is $H_{a}\oplus H_{b}$.
The point is that the vacuum operator on each connected worldsheet 
in
$\mathcal{P}$ can be interpreted as an element of 
\beg{ehom}{\bigotimes H\otimes Hom(H_{a}, H_{a}).}
Here $Hom$ stands, for example, for bounded homomorphisms. There is,
of course, an additional condition that the element \rref{ehom}
be trace class, but it is nevetheless clear that endomorphisms
of two Hilbert spaces can be added and that the sum of trace class
endomorphisms is trace class. It is then easily verified that such
choice of elements extends to a full structure of an $D$-brane module.

\vspace{5mm}
Fixing $1$-dimensional anomaly $\alpha$ ($\ct$-central extension) on $P$,
we now define the $D$-brane isomorphism category $C_{\alpha}(H)$ as follows: the
objects are finitely generated $D$-brane modules $H_{a}$, and morphisms
are isomorphisms of $D$-brane modules. This is a symmetric monoidal
category with respect to the direct sum just introduced. Its classifying space
is therefore an $E_{\infty}$-space, and the corresponding generalized
cohomology theory $D(H)$ is what we may call $D$-brane cohomology,
``classifying'' stable $D$-brane modules in a manner analogous to Moore-Segal's
theorem for $2$-dimensional TQFT (\cite{ms}).

\vspace{3mm}
It is appropriate to comment on how one would approach the construction
of a state space of string stretched between two $D$-brane modules, i.e.
how to use $D$-brane modules to get an actual $D$-brane category.
We remark, however, that we do not have that question 
solved mathematically. The problem is mainly the topology, or Hilbert space structure:
in closed CFT, Hilbert space structure comes from a combination of
a symmetric bilinear form got by a thin annulus with two inbound
(or outbound) boundary components, and real structure (see \cite{hk}
for a discussion). Similar ingredients can be used to give a Hilbert
structure for the state space $H_{aa}$ (which we are yet to construct).
However, no such natural Hilbert structure seems to be present on
$H_{ab}$ (just as there isn't on the non-unit labels of RCFT), because
these modules are not necessarily contragredient to themselves.

\vspace{3mm}
Nevertheless, the outline the construction of $H_{ab}$ is as follows:
consider all possible disk worldsheets $Q$ with two inbound parametrization components
and one outbound parametrization component. One of the inbound 
parametrization components
is labelled $a$ and the other one $b$. The other inbound parametrization
component is labelled $ab$, and its inbound point abuts the outbound point
of the parametrization component labelled $a$, while its outbound point
abuts the outbound point of the outbound parametrization component.
Now an element of $H_{ab}$ is a collection of trace class elements
$$f_{Q}\in H_{a}^{*}\hat{\otimes} H_{b},$$
one for each worldsheet $Q$ as above. The axiom is that whenever any number
of the
worldsheets $Q$ are glued to any number of worldsheets in $\mathcal{P}$ 
(preserving labels) and trace
of vacua taken (with $f_{Q}$ taken as vacuum on $Q$), then the resulting 
element coincides (via the canonical isomorphisms supplied by SLCMC) with
the element obtained by similar (but different) gluing which produces
an isomorphic worldsheet. It can be shown that state spaces constructed
in such way (``relative homomorphisms'') do indeed obey axioms of open
string CFT in some weak sense, but additional work is needed to
explain topology, and the trace class condition.

\vspace{5mm}
Finally, we remark that at least in one situation, there is
a good candidate at least for the Hilbert space $H_{aa}$. This occurs
when $H_{a}$ obeys the $D$-brane category axioms already, with open
label set $K^{\prime}=\{a\}$. Note that this is not such an
unreasonable
assumption, since the SLCMC testing these conditions {\em contains}
the partial SLCMC $\mathcal{P}$, and there might be a general reason
why the stronger condition should occur (see examples below). 
In that case, we call $a$ an {\em elementary $D$-brane}, and
we can set 
$$H_{aa}:=H_{a}.$$
Of course, it is not guaranteed that this choice will agree with
the one made in the previous paragraph (if it does, we call it
a {\em simple $D$-brane}). However, note that even starting from
simple $D$-branes $a$, we easily obtain $D$-branes which are not
elementary: simply take the direct sum $na$ of $n$ copies of $a$.


\vspace{3mm}
The formalism of SLCMC's is very powerful, much more so than
operads. However, in order to provide a more concrete
taste of our present construction, we note that if we restrict
attention to genus $0$ worldsheets with one outbound closed boundary
component and no open string component or no closed string
outbound boundary component per connected component, we
can describe the type of module structure we are considering
in terms of operads. Concretely, let $\mathcal{C}$ be
the operad of connected
closed string genus $0$ worldsheets with one outbound
boundary component. Now consider the space of all connected
worldsheets in $\mathcal{P}$ of genus $0$ and with no outbound
closed string boundary component. Recall that according
to the definition of $\mathcal{P}$, in addition
to the closed parametrization components,
there is one additional boundary component containing
two open parametrization components, one inbound and one outbounds.
Let $\mathcal{D}(n)$ be the
space of all such worldsheets with precisely $n$ inbound closed
string boundary components. We may ask what algebraic structure
is present on the space $\mathcal{D}$.

\vspace{3mm}
The answer is that there are two structures which commute in the
appropriate sense: one is a {\em right} $\mathcal{C}$-algebra
structure 
\beg{eright}{\mathcal{D}(n)\times \mathcal{C}(k_1)\times...\times
\mathcal{C}(k_n)\r\mathcal{D}(k_1+...+k_n).}
The other structure, however, is that of a monoid 
\beg{emon}{\mathcal{D}(m)\times\mathcal{D}(n)\r \mathcal{D}(m+n).}

It is, of course, possible to describe explicitly how the structures
\rref{eright}, \rref{emon} should commute, but it is more convenient
to express that in terms of monads: If $C$ is the monad associated
with the operad $\mathcal{C}$, then the same construction applied
to $\mathcal{D}$ gives a functor $D$ with is a right $C$-functor
from spaces into topological monoids (recall that right $C$-function
means that we have a natural transformation $DC\r D$ satisfying the
obvious axioms). 

\vspace{3mm}
Every time we have such $\mathcal{D}$, which we might call right
monoid $\mathcal{C}$-algebra, and a $\mathcal{C}$-algebra $R$,
we get a notion of an $R$-module with respect to $\mathcal{C},\mathcal{D}$.
The standard example of $\mathcal{D}$ is
\beg{estand}{\mathcal{D}(n)=\mathcal{C}(n+1),}
in which case we obtain what is known as $R$-modules via $\mathcal{C}$
(or $R,\mathcal{C}$-modules). However, the example of $\mathcal{D}$
given above shows there are important examples which do not arise
by way of \rref{estand}. It is interesting that such discussion doesn't
seem to have been previously made in the literature.	

Now if we restrict the structure $D$-brane module $H_{a}$ to the
very special type of genus $0$ worldsheets just discussed, we can
simply say that $H_{a}$ is a module (in the Hilbert sense)
of $H$ with respect to $\mathcal{C}$,
$\mathcal{D}$.


\section{Conformal field theory and string theory}
\label{s4a}

\vspace{3mm}

We will now consider the relationship between conformal field theory and
string theory, and the way it reflects on our investigation.
As a standing reference on string concepts, we
recommend one of the standard textbooks on the subject, e.g. \cite{gsw}
or \cite{pol}. One added feature of string theory is, of course,
supersymmetry, but we shall see soon that this turns out not to be the only 
complication. 

\vspace{3mm}
We will, therefore, begin our discussion with bosonic string theory
(superstrings will enter later). The essential point of string 
quantization is that conformal field theory quantizes parametrized strings,
while physical strings should be unparametrized. Now to pass from parametrized
strings to unparametrized, one needs a way to quantize the complex structure.
This problem is analogous to gauge fixing in gauge theory. 
In fact, this is more than just an analogy: from a strictly worldsheet point
of view, conformal field theory is indeed a $2$-dimensional quantum
field theory satisfying Schwinger axioms, and can be viewed as a gauge theory
in a certain sense; however, we do not need to pursue this here.
The important point is that in string
theory, complex structure gauge is
in fact needed to produce a consistent theory: conformal
field theory is anomalous and, in Minkowski space, contains states of negative norm.

\vspace{3mm}
The modern approach to gauge fixing in gauge theory, and to string quantization, 
is through
Fadeev-Popov ghosts and BRST cohomology. In the string theory case, we start
with a CFT $H_{m}$, the ({\em matter CFT}). In this case, BRST cohomology is
essentially a semi-infinite version of Lie algebra cohomology of the 
complexified Witt algebra (viewed as a ``Lie algebra of the semigroup of annuli'')
with coefficients in $H_{m}$. To be precise about this, we must describe
the semiinfinite analogue of the complex $\Lambda(g)$ for a Lie algebra $g$
where $g$ is the Witt algebra. As it turns out, this semiinfinite Lie complex
is also a CFT which is denoted as $H_{gh}$ and called the Fadeev-Popov
ghost CFT. A mathematical description is outlined in \cite{scft}, and
given in more detail in
\cite{spin}. In the chiral CFT case, $H_{gh}$ is a Hilbert completion
(with a chose Hilbert structure) of
\begin{equation}
\label{estring+}
\Lambda(b_{n}|n<0)\otimes\Lambda(c_{n},n\leq 0).
\end{equation}
In the physical case, both chiralities are present, and $H_{gh}$ is
a Hilbert completion of the tensor product of \rref{estring+}
with its complex conjugate. To understand why this is a semiinfinite
Lie complex of the Witt algebra, we write the generators of \rref{estring+}
as
\begin{equation}
\label{estring++}
\begin{array}{l}
b_{n}=z^{-n+1}\frac{d}{dz}\\
c_{n}=z^{-n-2}(dz)^{2}.
\end{array}
\end{equation}
So, the $b_{n}$'s are vector fields on $S^{1}$ (elements of the Witt algebra)
and the $c_{n}$'s are dual to the $b_{-n}$'s. An ordinary Lie complex would
be the exterior algebra on the duals $c_{n}, n\in\Z$. However, a special
feature of CFT is a choice of vacuum which allows us, even before gauge
fixing, to define correlation functions which are finite, albeit anomalous:
this is the mathematical structure known as Segal-type CFT, which we have
axiomatized in \cite{hk} and here. However, this choice of vacuum of $H_{m}$ is what
prompts the ``semininfinite approach'', where the exterior generators of 
$H_{gh}$ are not all of the $c_{n}$'s, but half of the $c_{n}$'s and half
of the $b_{n}$'s, as in \rref{estring+}. Now it turns out that for our
further discussion, it will be important to know explicitly one part of
the Virasoro action on $H_{gh}$, namely the conformal weights, or eigenvectors
of $L_{0}$. One may guess that $b_{n}$, $c_{n}$ should be eigenvectors of conformal
weight $-n$, but it turns out that one must decrease the conformal weights of
the entire complex \rref{estring+} by $1$, so the correct conformal weight
of a monomial in the $b_{n}$'s and $c_{n}$'s is $-k-1$ where
$k$ is the sum of the subscripts of the $b_{n}$'s and $c_{n}$'s in the monomial.
In fact, it turns out that the vacuum of the ghost theory, i.e. the element
of $H_{gh}$ assigned to the unit disk, is 
\begin{equation}
\label{estring+++}
b_{-1}c_{0}.
\end{equation}
Now the ghost CFT has an anomaly which is described by a $1$-dimensional
modular functor $L$ which has central charge $-26$ in the chiral
case (see \cite{scft}, \cite{spin}) and $(-26,-26)$ in the physical
case. (In the chiral case, there is an additional complication
that $L$ must be considered a super-modular
functor, see \cite{spin} and Section \ref{s5} below.)
A CFT $H_{m}$ is called {\em critical} if it has anomaly described by
the $1$-dimensional modular functor $L^{\otimes -1}$. The $26$'th
power of the $1$-dimensional free bosonic CFT described (briefly) in
the next section is critical in the physical sense (with both chiralities).

\vspace{3mm}
Now for a critical CFT $H_{m}$, there is a certain differential $Q$ 
(called BRST differential) on the (non-anomalous) CFT
\begin{equation}
\label{estringi}
\mathbb{H}=H_{m}\hat{\otimes} H_{gh}.
\end{equation}
In the chiral case, one has explicitly
\begin{equation}
\label{estringii}
Q=\cform{\sum}{r\in\Z}{}L_{r}^{H_{m}}c_{-r}-
\frac{1}{2}\cform{\sum}{r,s\in\Z}{}(r-s):c_{-r}c_{-s}b_{r+s}:-c_{0}
\end{equation}
(see \cite{dh}, formula (4.59)). Here $L_{r}^{m}$ are the Virasoro generators
acting on $H_{m}$, and $c_{n},b_{n}$, $n\in\Z$ are now understood as operators
on $H_{gh}$ in the standard way (see \cite{dh}). In the non-chiral case,
one must add to \rref{estringi} its complex conjugate. $Q$ is a differential,
which means that
\begin{equation}
\label{estringiii}
QQ=0.
\end{equation}
The cohomological dimension is called the ghost number. The $c_{n}$'s have ghost
number $1$, the $b_{n}$ have ghost number $-1$, so the ghost number degree of
$Q$ is $+1$. We shall fix the ghost number as an algebra grading, so $1$ has ghost
number $0$, but other conventions also exist.

What is even more interesting than \rref{estringiii}, 
however, is that $Q$ turns $\mathbb{H}$ into
a ``differential graded CFT''. If we use the usual notation where we write
for a CFT, as a lax morphism of SLCMC's, 
$$X\mapsto U_{X},$$
then we may define a differential graded CFT by the relation
\begin{equation}
\label{estringiv}
\sum (1\otimes... \otimes Q\otimes ... 1)U_{X}=0,
\end{equation}
with the correct sign convention. For simplicity, we assumed in \rref{estringiv}
that all boundary components of $X$ are outbound, the adjoint operator $Q^{*}$
is used on inbound boundary components. Physically, \rref{estringiv} is due
to the fact that $Q$ is a conserved charge corresponding to the Noether 
current of a supersymmetry (called BRST symmetry) of the Lagrangian of $\mathbb{H}$
(see \cite{dh}).

\vspace{3mm}
In any case, we now see that the BRST cohomology
\begin{equation}
\label{estringv}
H=H^{*}(\mathbb{H},Q)
\end{equation}
is a non-anomalous CFT. Infinitesimally, this implies that
$$[Q,L_{n}]=0$$
where $L_{n}$ are the standard Virasoro algebra (in our case in fact Witt algebra)
generators. However, more is true. In fact, one has
$$L_{n}=[Q,b_{n}],$$
so
\begin{equation}
\label{estringvi}
Qx=0\;\Rightarrow\; L_{n}x\in Im(Q).
\end{equation}
Because of 
\rref{estringvi}, $L_{n}$ actually act trivially on $H$, so $H$ is in fact
a TQFT (which means that $U_{X}$ only depends on the topological type of $X$).
Therefore, our machinery would certainly seem to apply to $H$, in fact so
would that of Moore-Segal \cite{ms}. There are, however, two difficulties.

\vspace{3mm}
First of all, $H_{m}$ may not actually be a CFT as we
defined it because of convergence problems. For example, when $H_{m}$
is the free bosonic CFT on the $(25,1)$-dimensional Minkowski space,
the inner product on the space $H_{m}$ is indefinite, so this space
cannot be Hilbert-completed with respect to its inner product. This
is more than a technical difficulty: in physical language, this is the
cause of the $1$-loop divergence of bosonic string theory. In our language,
this means that the state space of our would-be TQFT is infinite-dimensional,
so $U_{E}$ for an elliptic curve $E$ is infinity, or more precisely undefined.
So there isn't, in fact, any variant of the $(25,1)$-dimensional
free bosonic CFT for which the machinery outlined above would work mathematically
and produce a true TQFT.

\vspace{3mm}
In physics, this is an argument why the free bosonic string theory is not
physical, and one must consider superstring theory. Our definition of
CFT works on the SLCMC of superconformal surfaces, but the convergence
problems persist, i.e. again, for the free $(9,1)$-dimensional super-CFT,
the BRST cohomology would be TQFT is infinite-dimensional. Physicists
argue that the (infinite) even and odd parts of the TQFT are ``of equal dimension'',
and thus the $1$-loop amplitude vanishes (a part of the ``non-renormalization
theorem''). However, we do not know how to make this precise mathematically.

\vspace{3mm}
There is another, more interesting caveat, namely that $H$ is actually not
exactly the object one wants to consider as the physical spectrum of string
theory. Working, for simplicity, in the bosonic case, one usually restricts
to states of ghost number $0$, which, at least in the free case, is isomorphic
to the quotient $H_{0}$ of the
submodule $Z_{0}\subset H_{m}$ of states $x\in H_{m}$ satisfying
$L_{n}x=0$ for $n>0$ and $L_{0}x=x$, by the submodule $B_{0}$ of
states of null norm (the Goddard-Thorn no ghost theorem). In bosonic string
theory, the vacuum of $H_{m}$ is in $H_{0}$, but this is a tachyon,
which is not the vacuum of $H$: the vacuum in $H$ is $b_{-1}$, as
remarked above, and has
ghost number $-1$. In superstring theory, the tachyon is factored out
by so called GSO projection, while the vacuum of course persists, and has
also ghost number $-1$, but a different name, due to the different structure
of the theory, which we have no time to discuss here.

\vspace{3mm}
One may ask what mathematical structure there is on $H_{0}$ itself.
Here the answer depends strongly on whether we work chirally or not.
In the chiral case, Borcherds \cite{borch} noticed that $H_{0}$ is a
Lie algebra. The Lie algebra structure comes from $[u,v]=u_{0}v$
where $u_{0}$ is the residue of the vertex operator $Y(u,z)$.
The Jacobi identity follows immediately from vertex operator algebra
Jacobi identity. From CFT point of view, this operation is analogous
to the Lie bracket in Batalin-Vilikovisky algebras.

\vspace{3mm}
However, when both chiralities are present 
(which is the case we are interested in),
the rabbit hole goes deeper than that. First note that
the CFT vertex operator is not holomorphic, and curve integrals do not 
seem to be the right operations to consider. Instead, elements of $Z_{0}$
are operator-valued $(1,1)$-forms, and therefore can be naturally integrated 
over worldsheets. Indeed, one can see that
integration of an element of $Z_{0}$ over worldsheets
produces an {\em infinitesimal deformation} of CFT. Elements of $B_{0}$
also deform the CFT, but only by a gauge transformation, so elements of $H_{0}$
give rise to
infinitesimal deformations of
string theory. We may therefore (despite potentially serious convergence problems)
wish to consider a {\em moduli space} $M$ of string theories, to which $H_{0}$
is a tangent space at one point. In fact, points of the curved space $M$ should
be the true states of string theory, while the points of the tangent
space $H_{0}$ are only an approximation. In the physical theory,
one conjectures that the space $M$ contains all of the
$5$ original superstring theories as states, and a continuum of states
in between. As seen even by studying the basic example of toroidal spacetime,
some states in $M$ differ only by ``boundary conditions on open strings'',
and such conditions are called $D$-branes. When there is a well defined
spacetime manifold $X$, $D$-branes as a rule
are associated with submanifolds of $X$
with some additional structure. These, however, are classical and not quantum
objects (cf. Polchinski \cite{pol}), so that approach also has its
drawbacks. While rigorous 
mathematical attempts to define and investigate
$D$-branes from the manifold point of view have (with some success) also been made
in the literature
exist,
the ``tangent'' CFT approximation which we consider here is, in some sense, 
complementary. Finding a mathematical theory which would unify both points of view
is an even much more complex task, which we do not undertake here.

\vspace{5mm}

\section{An example: The $1$-dimensional free scalar CFT}
\label{s4}

We shall now give the standard examples of $D$-branes in the
free bosonic CFT in dimension $1$ (which is the CFT description of
the linear $\sigma$-model). Unfortunately,
even for this most basic CFT, a mathematically rigorous description
of its convergence issues
is nowhere to be found in the literature. The best outline 
we know of is given in \cite{scft}.

A good first guess for the free (bosonic) field theory state space is,
analogously with the lattice theories (see \cite{hk})
\beg{efree}{H=L^2(\R,\C)\hat{\otimes}\hat{Sym}<z^{n},\overline{z}^{n}|n>0>.}
Here $L^{2}(\R,\C)$ denotes $L^2$-functions with respect to the
Gaussian measure. The quantum number associated with this space
is the momentum. 

To be more precise, \rref{efree} should be a Heisenberg representation of
a certain infinite-dimensional Heisenberg group. To describe it, we
start with the topological vector space $V$
of all harmonic functions on $S^{1}$ or, more precisely, harmonic
functions on small open sets in $\C$ which contain $S^{1}$ and,
say, the topology of uniform convergence in an open set containing $S^{1}$. 
Thus, $V$ is topologically generated by the functions
$$z^{n},\overline{z}^{n},n\in\Z$$
and 
$$\ln||z||.$$
To define the Heisenberg group, we would like to find a $\C^{\times}$-valued 
cocycle
on $V$ which would be invariant under the action of $Diff^{+}(S^{1})$.
However, similarly as in the case of lattice theories, we do not
know any such cocycle. Instead, one considers the space $U$
of harmonic $\C$-valued functions on (an open domain containing) the
unit interval $I$. 
The point is that the harmonic functions on $I$ break up into
holomorphic and antiholomorphic parts; a topological basis of the
holomorphic part is given by the elements
$$z^{n}, n\in\Z,\; \ln(z),$$
and a topological basis of the antiholomorphic parts is given by their complex
conjugates. Therefore, the holomorphic and antiholomorphic parts $U_{+}$
and $U_{-}$ of $U$ have well defined winding numbers which can be added
to a total winding number; let $V^{\prime}\subset U$ be the set of functions
of total winding number $0$.
Then the map $exp(?)=e^{2\pi i?}$
gives a projection 
$$V^{\prime}\r V$$
whose kernel consists of the constant functions. Now to get the free field
theory, one proceeds analogously to lattice theories (see \cite{hk}),
specifying a cocycle on $U$. 
We shall specify separately cocycles on both $U_{+}$ and $U_{-}$. However, because
the integrality condition is replaced by equality of winding numbers on
$U_{+}$ and $U_{-}$, we have more freedom in choosing the cocycle. For
example, we can put, on both $U_{+}$ and $U_{-}$,
\beg{ecocyc1}{c(f,g)=exp
\frac{1}{2}(\int_{S^{1}}fdg-\Delta_{f}g(0)+\frac{1}{2}\Delta_{f}\Delta_{g}),}
(where $\Delta_{f}$ is the winding number). 
The effect of this is that if we apply the cocycle to lifts
of two harmonic functions $f$, $g$ on
a worldsheet to its universal cover, whose restriction to
boundary components are $f_{i}, g_{i}$ (as is done in \cite{hk} for
the lattice theories), the Greene formula implies that 
\beg{ecocyc}{\begin{array}{l}
c(f,g)=exp\frac{1}{4}(\cform{\sum}{i<j}{}(\Delta_{f_{i}}\Delta_{g_{j}}+\Delta_{g_{i}}
\Delta_{f_{j}})+\cform{\sum}{i=1}{n}\Delta_{f_{i}}\Delta_{g_{i}})=\\
exp\frac{1}{4}(\cform{\sum}{i=1}{n}\Delta_{f_{i}}\cform{\sum}{i=1}{n}\Delta_{g_{i}})=1.
\end{array}}
Thus, the situation is simpler than in the case of lattice theory.
Now similarly as in the case of lattice theory, the cocycle $c$ we have
constructed, when restricted to $V^{\prime}$, is trivial on the constant
functions, so we get a canonical map
\beg{econstmap}{\C\r \tilde{V^{\prime}}}
(where $\tilde{?}$ denotes Heisenberg group with respect to a given
cocycle). Similarly as in the case of lattice theories, in fact,
$c(f,g)=0$ for $f,g\in V^{\prime}$, $f$ constant, so the
subgroup \rref{econstmap} is normal, so the desired Heisenberg group
can be defined by 
\beg{eheis*}{\tilde{V}=\tilde{V^{\prime}}/\C.}
Then $H$ should be the Heisenberg representation of the right
{\em real form} of \rref{eheis*}. Note, however, that the above
construction comes with no obvious natural choice of real form.
Let us postpone the discussion of this issue, as we shall see it is
related to the convergence issues of the CFT.
Now, conformal field theory structure is specified as usual:
looking at the Heisenberg representation $H_{\partial X}$ of the central extension
$\tilde{V}_{\partial X}$
of the space $V_{\partial X}$ of harmonic functions on the boundary of a worldsheet $X$,
we have already constructed a canonical splitting of the pullback
of the central extension to the subspace $V_{X}$ of 
harmonic functions on $X$; we would like to define
the field theory operator associated with $X$
as the vector space of invariants of $H_{\partial X}$ with respect to $V_{X}$.

\vspace{3mm}
A usual ``density argument'' (cf. \cite{scft},
\cite{hk}, \cite{ps}) shows that the invariant vector space $H_{X}$
is always at most $1$-dimensional.
In more detail, if we denote by $Harm(X)$ the space of harmonic
functions on $X$ and by $Harm(\partial X)$ the space
of harmonic functions on a small neighborhood of $\partial X$,
and also by $D$ the unit disk, then, by restriction, we may form the
double coset space
\beg{eheis+}{Harm(X)\backslash Harm(\partial X)/\cform{\prod}{\partial}{}
Harm(D)}
(the product is over boundary components of $X$). Then \rref{eheis+}
is isomorphic to $H^{1}(\overline{X},\underline{Harm})\cong \C$
where $\underline{Harm}$ is the sheaf of harmonic functions and
$\overline{X}$ is the worldsheet obtained from $X$ by gluing 
unit disks on the boundary components. Since $H$ can be
interpreted as a (completed) space of functions on 
\beg{eheis++}{Harm(\partial X)/\cform{\prod}{\partial}{} Harm(D),}
the identification of \rref{eheis+} shows that only functions
on the orbits $\C$ have a chance to be  $Harm(X)$-fixed points.
However, studying further the constant functions in $Harm(X)$,
we see that only functions supported on $\{0\}\subset \C$ have
a chance of being fixed points.

These observations also point to a difficulty with a Hilbert
space model for $H$. What kind of reasonable Hilbert space functions
on $\R$ contain distributions supported on a single point?
Now recall the reason why the Hilbert space is not yet fixed:
we haven't fixed the real structure on $V$. One clue for such
real structure is that, from the desired interpretation of $H$ as
functions on \rref{eheis++},
$$A= Harm(D)\subset V$$
should be our ``Lagrangian subspace'' so that
$H=\widehat{Sym(A)}$, (cf. \cite{ps}, Section 9.5). Thus,
we can define the real structure on $V$ by specifying the inner
product on $A$. The choice enjoying the desired invariances
is
\beg{eheis+++}{\langle f,g\rangle=\int \overline{f}dg.}
But then this is only a semidefinite Hermitian product on $A$, where constants
have norm $0$! The above discussion shows that this is more
than a technical difficulty. To get the field operator $U_{X}$ 
to converge, we must take the inner product \rref{eheis+++}, which leads
to
$$H=\cform{\prod}{k\in\R}{}\widehat{Sym\langle z^n,\overline{z}^n|n>0
\rangle}$$
($\prod$ is the Cartesian product). Physically, the quantum
number $k$ is the momentum. Then the notion of ``Hilbert
product'' of copies of $H$ and ``trace'' must be adjusted.
For more details, see the Appendix. With these choices, convergence
of $U_{X}$ can be proven similarly as in \cite{hk} for lattice
theories (e.g. using boson-fermion correspondence
at $0$ momentum), so the free bosonic CFT is rigorous. 
This convergence
problem does not arise if we consider the $\sigma$-model on a compact
torus instead of flat Euclidean space.

For completeness, we note that we haven't discussed inbound
boundary components, and closed worldsheets. The former topic offers
no new phenomena and can be treated simply by reversing the sign of
the cocycle. Discussing closed worldsheets amounts really to
discussing in detail the anomaly, which is $H^{1}(X,\underline{Harm})$,
analogously to the fermionic case treated in \cite{spin}.

\vspace{3mm}
Now we want to give examples of simple elementary $D$-branes in the free CFT.
The above discussion shows that we would have to work in compact spacetime
(a torus) to make the examples fit the scheme proposed in Section \ref{s2}
literally. 
However, we elect instead to stick to flat spacetime $\R$, where the 
situation seems more fundamental. It must be then understood, however,
that the notion of $D$-brane module in this situation must also be
generalized in a way analogous to CFT (as discussed in the Appendix), to
solve the convergence issue.

\vspace{3mm}
Consider the $1/2$-disk $B$ consisting of elements of $D$ with
non-negative imaginary part. We consider $B$ an open string worldsheet
where the real boundary elements are the $D$-brane  component, and
the open string component is parametrized by the map $e^{\pi i t}$.
Then we can consider the space of all harmonic functions 
on the boundary of $B$ which
obey a suitable boundary condition on the $D$-brane component.
The boundary conditions allowable first of
all must be conformally invariant. The most obvious such condition
is that the derivative of the function in question be $0$ in
the direction of a certain vector $u\in S^{1}$, $Im(u)>0$ or $u=1$.
A priori, all of those conditions are allowable.
However, if we want to follow the methods we used above to describe
closed free CFT, additional conditions are needed. Namely, we need
the vector space of functions satisfying the condition to have
a central extension which is a Heisenberg group. Moreover, to
get consistency of open and closed CFT, we need the Heisenberg group
to be obtained by restriction of the cocycle \rref{ecocyc1}.
This means that the cocycle \rref{ecocyc1} or more 
precisely the
bilinear form by whose exponentiation the cocycle arises,
must be non-degenerate on the vector space of functions we are
allowing. This happens for the case $u=i$, although in order for
it to work, we must allow the harmonic function $\ln||z||$, which
has a logarithmic singularity on the $D$-brane boundary component of $B$!
Nevertheless, this case works, and gives the classical free open string theory
(von Neumann boundary condition).
The corresponding space of functions for $u=1$ is polarized canonically
(by the subspace of harmonic functions which extend harmonically to $B$,
with possibly logarithmic singularities on the $D$-brane component),
so we can take again the bosonic Fock space. This means that we
restrict the cocycle \rref{ecocyc1} to the vector space of
harmonic functions obeying the specified boundary condition, 
and take the Heisenberg representation. The field operator is
constructed in the same way as in the closed string case. However,
this time it is not immediately obvious that the space $H_{\partial X}$
for an open worldsheet $X$ is isomorphic to the suitably completed tensor product
of the state spaces of the individual parametrization components.
Nevertheless, this can be proven by sewing on copies of the half-disk
$B^{-}$ with opposite parametrization; the resulting gluing map 
proves that the 
space $H_{\partial X}$ is dual to the Hilbert tensor product of 
the individual copies of $H_{B^{-}}$, as needed.

\vspace{3mm}
We have therefore constructed classical open string theory, i.e.
the $1$-dimensional $D$-brane, but what about the $0$-dimensional
$D$-brane, i.e. the open free CFT with
a Dirichlet condition? 
(Note: in physics, one spacetime dimension is time, which is usually
not counted in the dimension of the $D$-brane; thus, instead of $1$- and
$0$-dimensional $D$-branes, we would speak of $D0$-brane and $D(-1)$-brane
or instanton.)
In any case,
in the language considered above, $u=i$ was the Von Neumann
condition, so $u=1$ should be the Dirichlet condition. However, in the
case of $u=1$ we are facing the same difficulties we had for general $u$,
namely that the functions satisfying the boundary condition do not
extend to a Heisenberg group using the cocycle \rref{ecocyc1}.
In order to get a Heisenberg group, we must find a way to ``get rid of
the constant functions'', and do so naturally enough to support a field
theory. It is not clear how to do that for general $u$, but for $u=1$
there is a special construction which eliminates the constants canonically:
take
$$V_{0}=
<f-\overline{f}|\text{$f$ is a harmonic function on $\partial B$}>.$$
It can be shown that, using the cocycle \rref{ecocyc1}, the Heisenberg representation
$H_{0}$ of $\tilde{V}_{0}$ is indeed an open string theory together with
the free closed string theory $H$. We see that the space $H_{0}$ misses
the momentum quantum number, but this doesn't mean the momentum is constant;
rather, as it turns out, the appropriate interpretation
is that the position is constant, and the space $H_{0}$ is the state
space of a $D$-brane of the free field theory $H$. We get a different elementary
$D$-brane for each choice of position.

\vspace{3mm}
We may ask if the $D$-branes at different positions we just described
are isomorphic as open CFT's of the $1$-dimensional closed free
bosonic CFT. The answer is that they are isomorphic, but not over
the identity automorphism of the
closed CFT. The free bosonic CFT has automorphisms coming from
translations: $D$-branes of positions $\mu$, $\lambda$ are
isomorphic over the closed CFT automorphism given by translation
by $\lambda-\mu$.

\vspace{5mm}

\section{CFT anomaly via $2$-vector spaces and elliptic cohomology}
\label{s5}

\vspace{3mm}
In this section, we give a new definition of modular functor
which generalizes the definition given in \cite{hk}. Consider a 
free finitely
generated lax module $\mathcal{M}$ over the lax semiring $\C_{2}$
(the category of finite-dimensional $\C$-vector spaces; it is convenient
to let the morphisms of $\C_{2}$ be all linear maps; thereby,
$\C_{2}$ is not a groupoid, and we have to use the version of lax algebra
theory which works over categories - see the appendix; 
of course, it is possible to consider
the subcategory $\C_{2}^{\times}$of $\C_{2}$ whose morphisms are isomorphisms).
Consider further $\C_{2}^{Hilb}$, the $\C_{2}$-algebra 
of Hilbert spaces with the Hilbert
tensor product. We put $\mathcal{M}^{Hilb}=\mathcal{M}\otimes_{\C_{2}}\C_{2}^{Hilb}$.
Now consider $H\in_{1}\mathcal{M}^{Hilb}$ (the symbol $\in_{1}$ means a map
of lax $\C_{2}$-modules $\C_{2}\r ?$; in our case,
this is the same thing as $H\in Obj(\mathcal{M}^{Hilb})$). We shall define two LCMC's $C(\mathcal{M})$
and $C(\mathcal{M},H)$ (underlying LCM of sets)
which are, in standard ways, extended into SLCMC's over
the Grothendieck category of finite-dimensional smooth complex manifolds.

\vspace{3mm}
The LCMC's are constructed as follows: the objects of $C(\mathcal{M})$ 
over the pair of finite sets $(S,T)$ are
$1$-elements of $\mathcal{M}^{\otimes S}\otimes \mathcal{M}^{*\otimes T}$; 
the morphisms are $2$-isomorphisms. 
Here $\mathcal{M}^{*}$ is the dual lax $\C_{2}$-module of $\mathcal{M}$,
whose objects are (lax) morphisms of lax $\C_{2}$-modules $\mathcal{M}\r
\C_{2}$ and morphisms are natural isomorphisms compatible with the
operations.
The gluing
maps are given by trace over $\C_{2}$, i.e. the evaluation morphism
$\mathcal{M}\otimes\mathcal{M}^{*}\r\C_{2}$. An object of $C(\mathcal{M},H)$ 
over $(S,T)$ consists
of an object $M$ of $C(\mathcal{M})$, and $2$-morphism 
$u:M\r_{2} H^{\hat{\otimes} S}\hat{\otimes}
H^{*\hat{\otimes} T}$ whose image consists of trace class elements:
(Choosing a basis of $\mathcal{M}$, a $1$-element of $\mathcal{M}^{Hilb}$
becomes a collection of Hilbert spaces, so $\hat{\otimes}$-powers
of $H$ are collections of $\hat{\otimes}$-powers; an element is
trace class if each of its components is trace class. For generalizations
beyond the trace class context, see Remarks in the Appendix.) 
Here $H^{*}\in_{1}\mathcal{M}^{*\; Hilb}$ is defined by
putting, for $V\in_{1}\mathcal{M}$, 
$H^{*}(V)=Hom_{\mathcal{M}^{Hilb}}(H,V)$.
Morphisms are commutative diagrams of the obvious sort.
To define gluing operations, note that 
$$u:M\r_{2}H^{\otimes(S+U)}\hat{\otimes} H^{*\hat{\otimes} (T+U)}$$
induces a $2$-morphism $tr(u):(1\otimes tr_{1})M\r_{2}H^{\hat{\otimes} S}
\hat{\otimes}H^{*\hat{\otimes} T}$ where $tr_{1}:\mathcal{M}^{\otimes U}
\otimes \mathcal{M}^{*\otimes U}\r\C_{2}$ is the evaluation morphism;
it is defined by using the canonical morphism  $tr_{2}:tr_{1}(H\otimes H^{*})\C$.

\vspace{3mm}
To define the corresponding SLCMC's, (which 
we denote by the same symbols),
as usual, it suffices to define sections over pairs of constant covering spaces
$(U\times S,U\times T)$ of a complex manifold $U$.
For defining $C(\mathcal{M})$,
we need a concept of a holomorphically varying $1$-element of $\mathcal{M}$.
To this end, we denote by $Hol(U,\C_{2})$ the lax commutative monoid
of finite-dimensional holomorphic bundles on $U$. Then the concept
we need is
$$M_{U}\in_{1}\mathcal{M}_{U}:=\mathcal{M}\otimes_{\C_{2}}Hol(U,\C_{2}).$$
This determines the SLCMC $C(\mathcal{M})$.
To define $C(\mathcal{M},H)$, note that $H$ does not depend on $U$, so in
the case of constant covering space described above, we simply need a
$$u_{U}:M_{U}\r_{2}H_{U}^{\hat{\otimes}S}
\hat{\otimes}H_{U}^{*\hat{\otimes}T}$$
where $H_{U}$ is the constant $H$-bundle on $U$, and the $2$-morphism means
a morphism of holomorphic bundles.

\vspace{3mm}
\noindent
{\bf Definition:} A {\em modular functor} on an SLCMC $C$ with labels
$\mathcal{M}$ is a (lax) morphism
of SLCMC's $\phi:C\r C(\mathcal{M})$. A CFT on $C$ with modular 
functor on labels
$\mathcal{M}$ with state space $H$
is a (lax) morphism of SLCMC's $\Phi:C\r C(\mathcal{M},H)$.

\vspace{3mm}
Now note that $C(?)$ is a $2$-functor from the $2$-category 
$\C_{2}-mod$ of lax $\C_{2}$-modules ($1$-morphisms are equivalences
of $\C_{2}$-modules, and $2$-morphisms are natural isomorphisms 
compatible with $\C_{2}$-module structure)
into the $2$-category of SLCMC's. Similarly, $C(?,?)$ is a $2$-functor
from the $2$-category $\C_{2}-mod_{*}$ of pairs $\mathcal{M},H$ where 
$\mathcal{
M}$ is a $\C_{2}$-module, and $H\in_{1}\mathcal{M}^{Hilb}$.
Here $1$-morphisms in $\C_{2}-mod$, $\C_{2}-mod_{*}$ are equivalences
of lax $\C_{2}$-modules, $2$-morphisms are natural isomorphisms compatible
with $\C_{2}$-module structure. (In $\C_{2}-mod_{*}$, $1$-morphisms
$(\mathcal{M},H)\r(\mathcal{N},K)$
are $1$-morphisms $\phi:\mathcal{M}\r
\mathcal{N}$ in $\C_{2}-mod$
together with a $2$-isomorphism 
$\lambda:\phi(H)\r K$; a $2$-morphism $\phi\r\psi$
is a $2$-morphism
in $\C_{2}-mod$ which induces an isomorphism $\phi(H)\r \psi(H)$
commuting with the $\lambda$'s.)

\vspace{3mm}
We use this to build a $2$-category $\Gamma$
of $CFT$'s as a ``comma $2$-category''. The objects are tuples 
$\mathcal{M},H,\Phi$ where $\Phi$ is a CFT on $C$ 
with labels $\mathcal{M}$ and state space $H$, $1$-morphisms are
tuples $\Phi,\Psi,f, \iota$ where $\Phi$, $\Psi$ are CFT's with labels
$\mathcal{M}$, $\mathcal{N}$ 
and state spaces $H$, $K$, $f$ is a 
$\C_{2}-mod_{*}$ - $1$-morphism $(\mathcal{M},H)\r(\mathcal{N},K)$
and $\iota$ is a natural isomorphism $f(M_{X})\r_{2}N_{X}$
where $M_{X},N_{X}$ are the $1$-elements of $\mathcal{M}$, $\mathcal{N}$
assigned to $X\in Obj\mathcal{C}$ by $\Phi$, $\Psi$
which commutes with SLCMC structure maps and the $u$'s assigned
by $\Phi$, $\Psi$ (we have used the notation of sections
over a point, but we mean this in the stack sense for sections
over any complex manifold $U$). $2$-morphisms $\Phi,\Psi,f,\iota\r\Phi,\Psi,g,\kappa$
are given by $2$-isomorphisms $f\r g$ in $\C_{2}-mod_{*}$ which
commute with $\iota$, $\kappa$ (hence the $u$'s).

\vspace{3mm}
Next, we shall show that $\Gamma$ is a symmetric monoidal $2$-category.
This means that
we have a lax $2$-functor $\oplus$ with
the same coherence $1$-isomorphisms as in a symmetric
monoidal category, but coherence diagrams commute up to $2$-cells;
the $2$-cells, in turn, are required to satisfy all commutations
valid for the trivial $2$-cells of coherence diagrams in an ordinary
symmetric monoidal category. 
Thus, the main point is to construct the $2$-functor $\oplus$. 
Suppose we have two objects $\mathcal{M},H,\Phi$ and $\mathcal{N},K,\Psi$
of $\Gamma$. Then their sum is $\mathcal{M}\oplus\mathcal{N},H\oplus K,\Phi\oplus\Psi$:
The first component is the direct sum in $\C_{2}-mod$. $H\oplus K$ is
the direct sum induced by that functor on $1$-morphisms. The symbol
$\Phi\oplus \Psi$, however, has to be defined explicitly. For simplicity,
we shall restrict to sections over a point. 
Then, the data which remains to be defined, for an object $X$
of the source SLCMC, is 
\beg{empn}{M\oplus N,} 
and
\beg{eupu}{u:M\oplus N\r_{2}(H\oplus K)^{\otimes S}\otimes (H\oplus K)^{*\otimes T}.}
Of this data, for $X$ connected,
\rref{empn} is, again, the direct sum induced by the direct sum
of $\C_{2}$-modules on $1$-morphisms, composed with the canonical
map
\beg{epush}{\begin{array}{l}\mathcal{M}^{\otimes S}\otimes\mathcal{M}^{*\otimes T}\oplus
\mathcal{N}^{\otimes S}\otimes \mathcal{N}^{*\otimes T}\\
\r
(\mathcal{M}\oplus\mathcal{N})^{\otimes S}\otimes
(\mathcal{M}\oplus\mathcal{N})^{*\otimes T}. 
\end{array}
}
Analogously, \rref{eupu} is given by the $\oplus$ of $\C_{2}$-modules
on $2$-morphisms, composed with \rref{epush}. For $X$ non-connected,
note that we are forced to define all data by applying the tensor
product to the data on connected components. This definition extends
to $1$-morphisms and $2$-morphisms in a standard way to produce
a symmetric monoidal $2$-category of CFT's $\Gamma$. 

Note that we may not always
wish to work with the whole $\Gamma$, but with some 
symmetric monoidal sub-$2$-category $\Lambda$; for example, we may take
direct $\oplus$-sums of copies of a given CFT.

\vspace{3mm}
Now there is an infinite loop space machine for
$2$-categories: for example,
Segal's machine. Segal's machine is supposed to
construct an $\mathcal{F}$-space, which is a functor from the
category $\mathcal{F}$ of finite sets with base point $*$
and based maps into spaces (alternately, one can think of this
as a category of partial maps); it is also required that the
functor (called $\mathcal{F}$-space) $B$ be {\em special}, which means
that the product map from $B(n)$ to the product of copies of $B(1)$
by the maps which send all numbers in $\{1,...,n\}$
except $i$ into the basepoint be an equivalence.

Now to produce a special $\mathcal{F}$-space from a symmetric
monoidal category $C$, simply consider the category $C(n)$
which is a category of diagrams, whose objects are tuples $(x_{T})$
of objects of $C$ indexed by non-empty subsets of $S$, together with isomorphisms 
\beg{esc}{\cform{\bigoplus}{i\in T}{}x_{\{i\}}\cong x_{T}}
Morphisms are commutative diagrams of the obvious kind (see \cite{sc}).
Now $C(?)$ is a functor from $\mathcal{F}$ into categories, so
applying the classifying space gives the requisite $\mathcal{F}$-space.
It is special by basic theorems about the homotopy
types of classifying spaces.

However, now note that the same definition
\rref{esc} in the case of a symmetric monoidal $2$-category $C$
gives a $2$-category $C(n)$. The only difference is that on $1$-morphisms,
we do not consider merely diagrams commutative on the nose, but up to $2$-cells
and $2$-morphisms are systems of $2$-cells which further commute with the
$2$-cells thus introduced. With that, however, $C(?)$ becomes
a (strict) functor from $\mathcal{F}$ into $2$-categories.

So, we are done if we can produce a functorial classifying space
construction $B_{2}$ on $2$-categories, and show that the 
$\mathcal{F}$-space $B_{2}C(?)$ is special. The latter is a straightforward
exercise which we omit. For the former, however, we remark that
to define a classifying space of a $2$-category $C$, we can first form 
the bar construction $B_{1}=B(Mor(C))$, i.e. the bar construction on
$2$-morphisms. However, if $C$ is lax, then $B_{1}$ is not a category,
but composition is defined with respect to a contractible operad
(without permutations). The operad $D$ is as follows:
the space $D(n)$ is the standard 
$(n-1)$-simplex and the composition is given by join:
$$D(k)\times D(n_{1})\times...\times D(n_{k})\r
D(n_{1})*...*D(n_{k})=D(n_{1}+...+n_{k}).$$
Nevertheless, it is well known that such 
$A_{\infty}$-categories
still have a classifying space functor (for example, one can ``rectify''
them by push-forward change of operads to the one point operad without
permutations, which encodes associativity).

\vspace{3mm}
Thus, we have produced a symmetric monoidal $2$-category of CFT's,
and an infinite loop space machine for such case. Therefore, we
have an infinite loop space $E$. This is related to the kind
of construction used in \cite{hk} to give a candidate for an elliptic-like
cohomology theory and seems like an improvement in the sense that it gives
a model for the {\em additive} infinite loop structure (a ``free'' construction
was used in \cite{hk}).

\vspace{3mm}
However, Baas-Dundas-Rognes \cite{rog} point out that this kind of
construction is naive. The problem is that there are not enough isomorphisms
of free lax $\C_{2}$-modules: they are essentially just permutation matrices
composed with diagonal matrices of line bundles. In \cite{rog}, 
a solution to this problem is proposed, conjecturally calculating the
algebraic $K$-theory of $\C_{2}$. 
The point is to consider, instead of invertible matrices of
finite-dimensional vector spaces, numerically invertible matrices,
which means that the corresponding matrix of dimensions of the entry
vector spaces is invertible.

Unfortunately, this approach does not seem satisfactory for the purposes
of CFT: along with an iso $f:\mathcal{M}\r\mathcal{N}$, we need to consider
also the inverse $\mathcal{M}^{*}\r \mathcal{N}^{*}$ of the adjoint morphism
$\mathcal{N}^{*}\r\mathcal{M}^{*}$; there is no candidate for such inverse
when $f$ is only numerically invertible.

\vspace{3mm}
Another clue that something else is needed is the following example of 
$bc$-systems, whose anomaly, it seems, can only be expressed by
considering ``modular functors with positive and negative parts''.

\vspace{3mm}
\noindent
{\bf Example:}
Consider the chiral $bc$-system of
$\Omega^{\alpha}$-forms, $\alpha\in\Z$
(see also Section \ref{s4a} above). The $bc$-system was first
considered mathematically by Segal \cite{scft}, but 
the observation that
the super-modular functor formalism is needed to capture its properties 
is due to
P. Deligne (\cite{spin}). In the case, the state space of the $bc$-system
is the ``fermionic Fock space''
\beg{edel1}{\mathcal{F}_{\alpha}=\hat{\Lambda}(H_{+}\oplus \overline{H}_{-})}
where $H=<z^{n}dz^{\alpha}|n\in\Z>$ and $H_{+}$ is, say, the subspace
$<z^{n}dz^{\alpha}|n\geq 0>$.
We select some real form to make this a positive definite Hilbert space
(cf. \cite{hk}, Chapter 2). Then the modular functor is $1$-dimensional,
and is given by the
determinant line of $\Omega^{\alpha}X$, the space of holomorphic
$\alpha$-forms on a worldsheet $X$. The reason
why a super-modular functor is needed here is
that we are dealing with Grassmanianns, and
signs must be introduced when permuting odd-degree
variables for the CFT to be consistent;
no such signs, however, occur in CFT's with $1$-dimensional
anomaly as considered above (see \cite{spin} for more details).

\vspace{3mm}
Thus, it seems that $\C_{2}$ in our definition of modular functors and
CFT's should be replaced by some sort of ``group completion'' which
would involve $\Z/2$-graded vector spaces. A candidate for such
construction is given in the next section, although we will see that 
this comes at the price of substantially increasing technical
difficulty.

\vspace{5mm}
\section{The group completion of $\C_{2}$.}
\label{s6a}

As argued above,
it would be desirable to have a group completion $\hat{\C}_{2}$
of $\C_{2}$ over which we could do the analogues of all of our constructions
as suggested by Baas-Dundas-Rognes \cite{rog}. In this Section, we
propose such construction. However, as also pointed out in \cite{rog},
any such construction is necessarily accompanied by sustantial difficulties.
The first problem is to even define what we mean by ``group completion''.
It is easy to show that for a lax commutative ring $R$, $BR$ is always
an Eilenberg-MacLane space (and hence cannot be used for our purposes),
but there is strong evidence that a large class of weaker categorical
notions of ``weakly group-complete'' lax commutative semirings suffer
from the same problem \cite{thomason}.

\vspace{3mm}
We take an alternate approach of introducing topology into the picture.
This means, we construct a model of a topolocial 
lax commutative semiring $\hat{\mathcal{C}}_{2}$ where
there is an object $-1$ so that $1\oplus (-1)$ is in the same connected
component of $0$. While this approach does seem to lead to a viable
definition, one must overcome a variety of technical difficulties
caused by the additional topology.

\vspace{3mm}
The first issue is what is the appropriate $2$-category $TCat$ of topological
categories? The point is that requiring functors to be continuous on
objects appears to restrict too much the notion of equivalence
of topological categories, and 
consequently alter their lax colimits. To remedy this situation, we
define $1$-morphisms $\mathcal{C}\r\mathcal{D}$ in $TCat$ to be of
the form
\beg{egp1}{{\diagram\protect
\mathcal{C}^{\prime}\rto^{F} \dto_{G} &\mathcal{D}\\
\mathcal{C}&\protect \enddiagram}}
where $F$ is a continuous functor and $G$ is a {\em partition} which
we define as follows: A partition is given by a topological space $X$
and a continuous map
$$f:X\r Obj(\mathcal{C})$$
such that the topology on $Obj(\mathcal{C})$ is induced by $f$ (we work
in the category of weakly Hausdorff compactly generated topological spaces).
Then we have $Obj(\mathcal{C}^{\prime})=X$, $Mor(\mathcal{C}^{\prime})$ is
a pullback of the form
$$
\diagram
Mor(\mathcal{C}^{\prime})\rto\dto& Mor(\mathcal{C})\dto^{S\times T}\\
X\times X\rto^{f\times f}& Obj(\mathcal{C})\times Obj(\mathcal{C}).
\enddiagram
$$
Two functors $F_{1}$, $F_{2}$ as in \rref{egp1} are considered {\em equal}
if they coincide on a common partition, i.e. we have a commutative diagram
$$\diagram
&\mathcal{C}_{2}^{\prime}\dlto_{G_{2}}\drto^{F_{2}}&\\
\mathcal{C}&\mathcal{C}^{\prime}\uto_{H_{2}}\dto^{H_{1}}\rto^{F}&\mathcal{D}\\
&\mathcal{C}_{1}^{\prime}\ulto^{G_{1}}\urto_{F_{1}}&
\enddiagram
$$
where $H_{1},H_{2}$ are partitions. Composition is defined by pullback
in the usual way, using the following

\begin{lemma}
\label{lgp1}
If we have a pullback
\beg{egpi}{{\diagram
Y^{\prime}\rto^{j}\dto_{h}&X^{\prime}\dto^{f}\\
Y\rto^{g}& X
\enddiagram}}
in the category of compactly generated weakly Hausdorff spaces such
that $f$ induces the (compactly generated) topology on $X$, then
$h$ induces the (compactly generated) topology on $Y$.
\end{lemma}

\Proof
Suppose $V\subset Y$, $v\in (Y-V)\cap Cl(V)$, $h^{-1}(V)$ closed.
Then there exists $K$ compact where $v$ is a limit point of $K\cap V$.
So, we may replace $V$ by $K\cap V$ and assume $Cl(V)=K=Y$. Next,
let $Z=g(K)$, so $Z$ is compact.

\vspace{3mm}
\noindent
{\bf Case 1:} $g(V)\neq Z$. Then there exists $T\subset X^{\prime}$
compact, $T\cap f^{-1}(g(V))$ not closed. Consider the pullback
$$\diagram
T^{\prime}\rto\dto & T\dto\\
g^{-1}f(T)\rto & f(T).
\enddiagram
$$
Then $T^{\prime}$ is compact since $g$ is proper. But then 
$j(T^{\prime}\cap h^{-1}(V))=f^{-1}(g(V))\cap T$, so
$T^{\prime}\cap h^{-1}(V)$ cannot be closed in $T^{\prime}$ (since
$j|T^{\prime}$ is closed, $T^{\prime}$ being compact). This is a
contradiction.

\vspace{3mm}
\noindent
{\bf Case 2:} $g(V)=Z$. Then in particular, there exists $z^{\prime}\in V$,
$g(v)=g(z^{\prime})=:z$. Now we may assume that
\beg{egpl+}{\text{$v$ is a limit point of $V\cap g^{-1}(\{z\})$.}
}
Indeed, otherwise, since $K=Y$ is compact weakly Hausdorff, it is normal,
hence regular and there exist $U,W$ open in $K$, $U\cap W=\emptyset$,
$v\in U$, $Cl(V\cap g^{-1}(\{z\}))\subset W$. But then we may
replace $Y$ by $Y-V$ (and $X$ by $g(Y-V)$), and we are back to Case 1.
So we may assume \rref{egpl+}. But then we may replace $X$ by $\{z\}$
and $Y$ by $g^{-1}(\{z\})$. But then \rref{egpi} is a product, in
which case the statement of the Lemma is obviously true (a product
projection induces the topology on its target in the compactly generated
weakly Hausdorff category). Thus, we have a contradiction again.
\qed

Now $2$-morphisms in $TCat$ are defined as follows: we can assume we have two
$1$-morphisms $F_{1}, F_{2}$ given as
$$\diagram
\mathcal{C}^{\prime}\rto^{F_{1},F_{2}}\dto_{G}&\mathcal{D}\\
\mathcal{C} &
\enddiagram
$$
where $G$ is a partition. Then a $2$-morphism is given by a partition
$$\diagram 
\mathcal{C}^{\prime\prime}\rto^{G^{\prime}}&\mathcal{C}^{\prime}
\enddiagram
$$
and a continuous natural transformation
$$F_{1}G^{\prime}\r F_{2}G^{\prime}.$$
Again, two $2$-morphisms are identified if they coincide after pullback
via a partition, similarly as above.

\vspace{3mm}
This completes the definition of the $2$-category $TCat$. This
$2$-category is defined in such a way that it has lax limits
defined in the same way as in $Cat$ \cite{fioret}.
(Lax limit is given as the category whose objects and morphisms
are lax cones from a point or an arrow to a diagram with the
topology induced from the product; by ``lax'', we always mean
``up to coherences which are iso''.)

\vspace{3mm}
Next, one may discuss lax monads in $TCat$, which, in our definition,
are lax functors $C:TCat\r TCat$ with lax natural transformations
$$\mu:CC\r C,\;\eta:Id\r C$$
which are associative and unital up to coherence isos with commutative
coherence diagram same as for lax monoids. For a lax monad $C$ in
$TCat$ we then have a category of lax $C$-algebras whose objects
are objects $M$ of $TCat$ together with a functor
$$\theta:CM\r C$$
satisfying associativity and unitality up to coherence isos with
commutative coherence diagrams of the same
form as those for categories with lax action of
a lax monoid.

Then lax algebras over a lax monad in $TCat$ form a $2$-category
which has lax limits created by the forgetful functor to $TCat$.
We may be interested in lax algebras over a strict monad, for example
the monad associated with a theory $T$. One example of a lax monad
whose lax algebras we are interested in is gotten from a $2$-theory
$(\Theta,T)$ and a lax $T$-algebra $I$. Then we can define a lax
monad $C_{\Theta,I}$ not over $TCat$, but over the category
$Tcat^{I^{k}}$ of strict functors $I^{k}\r TCat$. In effect,
$C_{\Theta,I}(X)$ has
$$C_{\Theta,I}(X)_{i}=\coprod \Theta(m)((\gamma_{1},...\gamma_{p});\gamma)$$
where the coproduct is indexed over $(j_{1},...j_{m})\in I$,
$\gamma\in T(m)^{k}$, $\gamma(j_{1},...j_{m})=i$, $\gamma_{1},...\gamma_{p}
\in T(m)^{k}$. Then lax $C_{\Theta,I}$-algebras are precisely lax algebras
over $(\Theta,T)$ with underlying lax $T$-algebra $I$.

\vspace{3mm}
We are now ready to describe a topological lax semiring $\hat{\C}_{2}$ with
an object $-1$ such that $1\oplus(-1)$ is in the same connected component
as $0$. First consider the lax semiring $s\C_{2}$ of pairs $(V_{+},V_{-})$,
$V_{+},V_{-}\in Obj(\C_{2})$ with the lax $\C_{2}$-module structure
given by $\C_{2}\oplus\C_{2}$ and multiplication
$$(V_+,V_-)\otimes(W_+,W_-)=(V_+\otimes W_+\oplus V_-\otimes W_-,
V_+\otimes W_-\oplus V_-\otimes W_+).$$
Now in $s\C_{2}$, consider the full subcategory $J$ on pairs $(V_+,V_-)$
where $dim(V_+)=dim(V_-)$. Then $J$ is a lax $s\C_{2}$-module, and
$J\oplus \C_{2}$ is a lax commutative $\C_{2}$-algebra with a lax
commutative $\C_{2}$-algebra morphisms $J\oplus C_{2}\r \C_{2}$
(an augmentation) and $J\oplus \C_{2}\r s\C_{2}$ (the inclusion).
Thus, we have a lax simplicial commutative $\C_{2}$-algebra
(=lax functor $\Delta^{Op}\r \text{lax commutative $\C_{2}$-algebras}$)
\beg{egpp+}{B_{\C_{2}}(\C_{2},J\oplus \C_{2},s\C_{2}).}
We propose $\hat{\C}_{2}$ to be the realization of \rref{egpp+}.
This needs some explaining, namely we must define realization.
We shall describe a lax realization functor 
$$\mathcal{A}_{\cdot}\r|\mathcal{A}|$$
from lax simplicial
commutative $\C_{2}$-algebras to topological commutative $\C_{2}$-algebras
(i.e. commutative $\C_{2}$-algebras in $TCat$). 
We want to mimic as closely as possible the strict construction. This
means that we will define \rref{egpp+}
as the lax simplicial realization in the $2$-category
of lax $\C_{2}$-modules, which we must define. First, let, for a space $X$,
$\C_{2}X$ be the free lax $\C_{2}$-module on $X$ (objects and morphisms
are finite formal
linear combinations with coefficients in objects and morphisms of $\C_{2}$, 
and the topology is induced from the topologies of finite powers of $X$).
We want the realization $|\mathcal{A}|$ to be the lax coequalizer of
\beg{egpp++}{
\cform{\oplus}{m,n}{}(\C_{2}\Delta_{m}\otimes \mathcal{A}_{n})
\begin{array}{c}\r\\[-.5ex]\r\end{array}\cform{\oplus}{n}{}
(\C_{2}\Delta_{n}\otimes \mathcal{A}_{n})
}
(the arrows are the usual two arrows coming from lax simplicial structure,
$\oplus$, $\otimes$ are over $\C_{2}$). To construct the lax coequalizer
\rref{egpp++}, we can take the objects of
\beg{egpp+++}{\cform{\oplus}{m,n}{}(\C_{2}\Delta_{m}\otimes \mathcal{A}_{n})
\oplus
\cform{\oplus}{n}{}
(\C_{2}\Delta_{n}\otimes \mathcal{A}_{n}).
}
To get morphisms, we take the morphisms of \rref{egpp+++}, and adjoin
isomorphisms between all source and target objects of the 
arrows in \rref{egpp++}. Take the free topological category
spanned by these morphisms, modulo the obvious commutative diagrams
required. This gives us a category with the lax $\C_{2}$-module
\rref{egpp+++} as a subcategory. The free construction we must then perform
is applying the {\em strict} left adjoint to the 
forgetful functor from the category of lax $\C_{2}$ modules
with lax submodule \rref{egpp+++} on the same set of objects (taking only
functors which are identity on objects) to
the category of categories with subcategory \rref{egpp+++} on the same set of
objects (taking only functors which are identity on
objects). As usual, the functors are strict because objects 
and coherences are
already specified. 

This completes the construction of the lax simplicial realization \rref{egpp+}.
One must still prove that this is a lax $\C_{2}$-algebra, but this is
accomplished analogously as in the strict case, using the shuffle map
(and the morphism definition \rref{egp1} to assure continuity).

\vspace{3mm}
Now topological SLCMC's $C(\mathcal{M})$, $C(\mathcal{M},H)$
for a finitely generated free topological lax $\hat{\C}_{2}$-module $\mathcal{M}$
are defined analogously as over $\C_{2}$. (Since the underlying stack of
covering spaces $I=Set$ does not change, LCMC's can be described as lax
algebras over a lax monad in $TCat$ as above, and therefore
stacks over a Grothendieck topology can be defined to be, as usual,
contravariant functors which take Grothendieck covers to lax limits.)

\vspace{3mm}
However, the topology would be of little use if we simply took for
our definition of modular functor a lax morphism of
SLCMC's from $\mathcal{C}$ to $C(\mathcal{M})$ (similarly for CFT's).
Instead, the corresponding ``derived notion'' is appropriate. This
means that we should consider lax morphisms of topological SLCMC's
\beg{egpp*}{
B(C_{\Theta,S},C_{\Theta,S},\mathcal{C})\r C(\mathcal{M})
}
where $\Theta$ denotes the $2$-theory of LCMC's, and $S$ the lax commutative
monoid of finite sets, as above. The left hand side is obtained by taking
the bar construction section-wise and then applying the lax left adjoint
to the forgetful functor from SLCMC's to pre-stacks of LCMC's.

It still remains to define a realization functor from lax simplicial
lax $C_{\Theta,S}$-algebras to topological lax $C_{\Theta,S}$-algebras.
Analogously as in the case of $B_{\C_{2}}$, however, we may simply lax-realize
in the $2$-category $(TCat)^{Set^{2}}$ (where it is easy to construct
lax colimits, cf. \cite{fioret}), and use the lax analogue of Milnor's
map $|A|\times |B|\r |A\times B|$ to obtain lax $C_{\Theta,S}$-algebra
structure on the realization. We omit the details.

\vspace{5mm}
\section{The general anomaly for open-closed CFT.}
\label{s7}

In this section, we shall apply the principles of
Section \ref{s5} to propose the a general definition of
open-closed CFT with both multiple $D$-branes and multi-dimensional
conformal anomaly (although without any group completion). 
We shall see, however, that this is necessarily
even much more complicated than what we have done in Section \ref{s5}.
We have already argued that neither the ``set of $D$-branes''
nor the ``set of labels'' should be sets. Rather, they should be
higher vector spaces. However, on a boundary component of the
worldsheet where several open parametrization components are
present, we need to take traces of ``sets of labels'' over
``sets of $D$-branes''. This suggests that our model of
``set of $D$-branes'' must be one categorical level above
our notion of ``set of labels''.

\vspace{3mm}
Therefore, we propose that the ``set of $D$-branes'' be a
$3$-vector space $\mathcal{A}$. When dealing with $3$-vector
spaces, note that they are $2$-categories. 
$3$-vector spaces are, by definition, $2$-lax modules
over the $2$-lax commutative semiring $\C_{3}$. 
We must, of course, define these notions. On generalizing
from lax to $2$-lax structures, we find it easiest to follow
the approach of \cite{hkv}. Let $T$ be a theory. Then let
$Th(T)$ be the free theory on $T$, with the canonical projection of
theories $\phi:Th(T)\r T$. Let $G$ be a groupoid with objects $Th(T)$
and one isomorphism $x\r y$ for every $x,y\in Th(T)$ which satisfy
$\phi(x)=\phi(y)$. Then $(Th(T),G)$ is a theory (strictly) enriched
over categories and a lax $T$-algebra is the same thing as a strict
$(Th(T),G)$-algebra enriched over categories. 

Now to go to the next level, consider the forgetful functor
$$U:\text{Theories enriched over groupoids} \r \mathcal{P}$$
where $\mathcal{P}$ is the (strict) category of pairs $(T,G)$
where $T$ is a theory, $G$ is a graph with objects $T$. Then
let $F$ be the left adjoint of $U$. Notice that $F$ is the
identity on objects $T$, so we may write
$$F(T,G)=(T,F(G)).$$
Now we have a map of theories enriched over groupoids:
$$\psi:(Th(T),F(G))\r (Th(T),G).$$
Therefore, we may create a $2$-category $(Th(T),F(G),H)$
by putting exactly one $2$-isomorphism between every $\alpha,\beta\in F(G)$
with $\psi(\alpha)=\psi(\beta)$. Then the $2$-category 
$(Th(T),F(G),H)$ is naturally a theory (strictly) enriched
over $2$-categories, and a {\em $2$-lax $T$-algebra} is
a $2$-category which is a strict $(Th(T),F(G),H)$-algebra
enriched over $2$-categories. (Obviously, one may proceed further
in the same way to define even higher laxness, but we shall not
need that here.)

One remark to be made is that theories, strictly speaking, do not
model universal algebras which are modelled on more than one set,
such as module over a ring (which is modelled over two sets).
However, algebras modelled over $k$ sets can be easily included
in the formalism by modifying the concept of theory to a category
with objects $\N^{k}$ (i.e. $k$-tuples of natural numbers) with
the axiom that for all $a,b\in \N^{k}$, $a+b$ is the categorical product
of $a,b$. All of our constructions generalize to this context.

\vspace{3mm}
Now to identify the categorical
levels with the levels we considered before, we will denote
objects of $3$-vector spaces by $\in_{0}$ and morphisms of
$3$-vector spaces by $\r_{0}$. Thus, a $0$-morphism
$\C_{3}\r\C_{3}$ (where $\C_{3}$ is the lax symmetric
monoidal category of $2$-vector spaces) is a $2$-vector space,
and the notations $\r_{1}$, $\r_{2}$ of such $2$-vector spaces
will coincide with the notations we used above.

\vspace{3mm}
Now given the $3$-vector space $\mathcal{A}$ (``the set of $D$-branes''),
we must introduce the ``set of labels'' for anomalies. The 
``set of closed labels'' will be, as before, a $2$-vector space, 
which we will denote by $\mathcal{C}$. The ``set of open labels''
will be an object of the form
$$\mathcal{O}\in_{0}\mathcal{A}\otimes_{\C_{3}}\mathcal{A}^{*}.$$
(We remark here that $\mathcal{A}^{*}$ for a $3$-vector space
$\mathcal{A}$ is defined analogously
as in the case of $2$-vector spaces.)
Now we would like to
define an SLCMC $C(\mathcal{A};\mc,\mathcal{O})$.
All our SLCMC's in this Section shall have two labels, $1$ and $m$ (closed
and open).
However, note that there is another subtlety we must provide for,
namely that the set $\Gamma$ of all possible {\em graphs} whose vertices are
open and closed parametrization and $D$-brane components and edges
describe their incidence relations with the obvious conditions (e.g.
all vertices have degree $2$ etc.) is itself an SLCMC, and in order
to correctly keep track of incidences on the boundary, we must
consider SLCMC's over $\Gamma$.

\vspace{3mm}
We shall only describe sections of $C(\mathcal{A};\mc,\mathcal{O})$
over a given object $G$ of $\Gamma$ over a single point, over four given
sets $S_{1,in}, S_{1,out}, S_{m,in}, S_{m,out}$
of inbound and outbound closed and open ``components''. Let $P$
denote the set of closed $D$-brane components of $G$.
Before making the definition, note that we have canonical dual $0$-morphisms
\beg{eco+}{{\diagram
\C_{3}\rto^{\eta}& \mathcal{A}\otimes_{\C_{3}}\mathcal{A}^{*}\rto^{\epsilon}&
\C_{3}.\enddiagram}}
(If no further discussion is made, \rref{eco+} requires a finiteness assumption about
$\mathcal{A}$.)
Their composition is a $2$-vector space which we shall denote by $tr_{0}\mathcal{A}$.
The set of sections of $C(\mathcal{A};\mc,\mathcal{O})$ are $1$-elements
\beg{eco1}{
\begin{array}{l}
M\in_{1} \cform{\bigotimes}{S_{1,in}}{}\mc^{*}\otimes
\cform{\bigotimes}{S_{1,out}}{}\mc \otimes\\[1ex]
\cform{\bigotimes}{P}{}tr_{0}\mathcal{A}\otimes
tr_{0,cyclic}(\cform{\bigotimes}{S_{m,in}}{}\mathcal{O}^{*}\otimes
\cform{\bigotimes}{S_{m,out}}{}\mathcal{O}).
\end{array}
}
Here the tensor products are over $\C_{2}$, and $tr_{0,cyclic}$ denotes
composition with the tensor product of the appropriate number of $\epsilon$'s;
note that although not explicitly written, the definition of $tr_{0,cyclic}$
makes use of all of the structure of $G$. Now in order to give \rref{eco1}
a structure of LCMC, one must show an appropriate gluing property, but this
is analogous to our discussion for closed CFT's.

\vspace{3mm}
Now let, as above, $\mathcal{B}$ be the SLCMC of closed-open worldsheet
with one closed and one open label. Then an open-closed CFT anomaly
(modular functor) is a map of SLCMC's over $\Gamma$
$$\mathcal{B}\r C(\mathcal{A};\mathcal{C},\mathcal{O}).$$
Now to define open-closed CFT, we must add the ``Hilbert spaces''. The
``closed Hilbert space'' is, as above,
$$H\in_{1}\mathcal{C}^{Hilb}.$$
The ``open Hilbert space'' is a $1$-morphism
$$K:\eta\r_{1}\mathcal{O}^{Hilb}$$
where $\eta$ is as in \rref{eco+}. We shall now define an SLCMC 
$C(\mathcal{A};\mathcal{O},\mc;H,K)$ over $\Gamma$. As above, we
shall specialize to sections over a single point and single object
of $\Gamma$, with the same notation as above. Then a section consists
of a section \rref{eco1} of $C(\mathcal{A};\mc,\mathcal{O})$ and
a $2$-morphism
$$\begin{array}{l}
M\r_{2} \cform{\hat{\bigotimes}}{S_{1,in}}{}
H^{*}
\hat{\otimes}
\cform{\hat{\bigotimes}}{S_{1,out}}{} H\hat{\otimes}
\otimes \cform{\bigotimes}{P}{}\eta_{tr_{0}\mathcal{A}}\\[1ex]
\hat{\otimes}tr_{1,cyclic}(\cform{\hat{\bigotimes}}{S_{m,in}}{}
K^{*}\hat{\otimes}\cform{\hat{\bigotimes}}{S_{m,out}}{}
K).
\end{array}
$$
Here $tr_{1,cyclic}$ is given by the structure of $2$-category, and
$\eta_{tr_{0}\mathcal{A}}$ is the canonical ``unit'' $1$-element
of $tr_{0}\mathcal{A}$. To be more precise, write, in \rref{eco+},
$$\eta(\C_{2})=\cform{\bigoplus}{i=1}{n}\mathcal{V}_{i}\otimes_{\C_2}\phi_{i},$$
so
$$tr_{0}\mathcal{A}=\cform{\bigoplus}{i=1}{n}\phi_{i}\mathcal{V}_{i}.$$
But then one can show
$$\phi_{i}\mathcal{V}_{j}=\delta_{i}^{j}\C_{2},$$
so we have
$$tr_{0}\mathcal{A}=\cform{\bigoplus}{i=1}{n}\C_{2},$$
and we can write
$$\eta_{tr_{0}\mathcal{A}}=\cform{\bigoplus}{i=1}{n}\C\in_{1}\cform{\bigoplus}{i=1}{n}
\C_{2}.$$
Of course, such discussion reveals the weaknesses of the higher vector
space formalism, and the desirability to really work, again, in a suitable
higher group completion. However, we do not work out that approach here.

\vspace{5mm}
\section{Appendix: Stacks of lax commutative monoids with cancellation}

\vspace{5mm}

To make this paper self-contained, we review here the basic definitions
\cite{hk} related to stacks of lax commutative monoids with cancellation
(SLCMC's).
We must begin by defining lax algebras. The formalism we use is theories
accoring to Lawvere, and their extension which we call $2$-theories.
Recall first that a {\em theory}
according to Lawvere \cite{law} is a category $T$ with objects
$\N$ (the set of all natural numbers $0,1,2,...$) such that
$n$ is the product of $n$ copies of $1$. Categories of
algebraic structures given by a set of operations and relations
on one set $X$ can be encoded by a theory, where $T_{n}=Hom(n,1)$
is the set of all $n$-ary operations of the algebraic structure
(including all possible compositions, repetitions of one
or more variables, etc.). 

\vspace{3mm}
\noindent
{\bf Definition:}
A {\em $2$-theory} consists of a 
natural number $k$, a theory $T$
and a (strict) contravariant functor $\Theta$ from $T$ to the
category of categories (and functors) with the following
properties. Let $T^{k}$ be a category with the same objects as $T$,
and $Hom_{T^{k}}(m,n)= Hom_{T}(m,n)^{\times k}$ (obvious composition). Then
$$Obj(\Theta(m))=\cform{\coprod}{n}{}Hom_{T^{k}}(m,n),$$
on morphisms, $\Theta$ is given by precomposition on $Obj(\Theta(m))$,
and
$\gamma\in Hom_{T^{k}}(m,n)$
is the product, in $\Theta(m)$, of the $n$-tuple
$\gamma_{1},...,\gamma_{n}\in Hom_{T^{k}}(m,1)$
with which it is identified by the fact that $T$ is a theory.
We also speak of a $2$-theory fibered over the theory $T$.

\vspace{3mm}
The example relevant to CFT is the $2$-theory of {\em commutative monoids
with cancellation}. $T$ is the theory of commutative monoids with an operation
$+$, and $k=2$. The $2$-theory $\Theta$ has three generating 
operations, addition (or disjoint union)
$+:X_{a,c}\times X_{b,d}\r X_{a+b,c+d},$
unit
$0\in X_{0,0}$
and cancellation (or gluing)
$\check{?}:X_{a+c,b+c}\r X_{a,b}.$
The axioms are commutativity, associativity and unitality for $+,0$,
transitivity for $\check{?}$
and distributivity of $\check{?}$ under $+$.

\vspace{3mm}
To get further, one needs to define algebras and lax algebras
over theories and $2$-theories. 
An algebra over a theory $T$ is a set $I$ together with,
for each $\gamma\in T_{n}$, a map $\overline{\gamma}:I^{\times n}\r I$, 
satisfying
appropriate axioms. These axioms can be written out explicitly,
but a quick way to encode them is to notice that 
for a set $I$, we have the {\em endomorphism theory} $End(I)$ where
$End(I)(n)=Map(I^{\times n},I)$, and we may simply say that a structure
of $T$-algebra on $I$ is given by a map of theories
$T\r End(I)$. 

\vspace{3mm}
To define an algebra over a $2$-theory $\Theta$ fibered over a theory
$T$, we must first have an algebra $I$ over the theory $T$ (the `indexing
theory'). This gives us, for $\gamma\in Hom_{T^{k}}(m,1)$,
a $k$-tuple of maps $\overline{\gamma}:I^{\times m}\r I$.
In an algebra over the $2$-theory, we have, in addition, a
map
\begin{equation}\label{exx}
X:I^{\times k}\r Sets.
\end{equation}
For a morphism in $\phi\in Mor(\Theta)$ 
from $(\gamma_{1},...,\gamma_{n})\in Hom_{T^k}(m,n)$
to $\gamma\in Hom_{T^k}(m,1)$, we have, for each choice $i_{1},...,i_{m}$
of elements of $I$, maps
\begin{equation}
\label{exx12}
\overline{\phi}:X(\overline{w_1}(i_{1},...,i_{m}))\times...
\times X(\overline{w_n}(i_{1},...,i_{m}))\r X(\overline{w}(i_{1},...,i_{m})),
\end{equation}
satisfying appropriate axioms. Once again, we can avoid writing them
down explicitly by defining the endomorphism $2$-theory.
Consider a set $I$ and a map
$$X:I^{k}\r Sets.$$
To such data there is assigned a $2$-theory $End(X)$ fibered over
the theory $End(I)$: let
$$\Theta(w;w_{1},...,w_{n})$$
consist of the set of all possible simultaneous choices of maps
\beg{ethsim}{X(w_{1}(i_{1},...,i_{m}))\times ...\times X(w_{n}(i_{1},...,i_{m}))
\r X(w(i_{1},...,i_{m}))}
where $i_{j}$ range over elements of $I$. A structure of
an {\em algebra} over the $2$-theory $\Theta$ fibered over $T$
is given by a morphism of $2$-theories
$$(\Theta, T)\r (End(X),End(I)).$$

\vspace{3mm}
A {\em lax algebra} over a theory is a category $I$, 
with maps
$\overline{\gamma}$ which are functors.
We do not, however, require that these maps define a strict
morphism from $T$ to the endomorphism theory of $I$. Instead,
this is only true up to certain natural isomorphisms, which
we call coherence isomorphisms, which in turn are required
to satisfy certain commutative diagrams, which are called coherence
diagrams.
This is, of course, always the case when defining lax algebras
of any kind. But now the benefit of introducing theories is
that the coherences and coherence diagrams always have the same shape. To be
more precise, recall that the notion of theory itself is an algebraic
structure which can be encoded by the sequence of sets $T(n)$,
and certain operations on these sets satisfying certain identities.
Denoting the set of operations defining
theories by $G$ (for `generators'), and identities by $R$
(for `relations'),
we observe that the set of coherence isomorphisms we
must require for lax $T$-algebras
is always in bijective correspondence with $G$, while the 
set of coherence diagrams needed is in bijective correspondence with $R$!

\vspace{3mm}
The concept of lax algebra over a $2$-theory is defined in a similar
fashion, but one important point is that one doesn't want
to consider the most general possible type of laxness (since that would
lead to a $3$-category). Rather, one starts with a lax algebra $I$ over
the indexing theory, and a {\em strict functor} 
$$X:I^{\times k}\r Categories; $$
appropriate
coherence isomorphisms and diagrams then follow in the same way as in
the case of lax algebras over a theory (are indexed by operations
and identities of $2$-theories interpreted as a `universal algebras' - see 
\cite{hk}).

\vspace{3mm}
The lax commutative monoid we most frequently consider is the
groupoid $S$ of finite sets and isomorphisms (the operation
is disjoint union). More generally, we
often consider a set of {\em labels} $K$ and the lax commutative
monoid of $S_K$ of sets $A$ labelled by $K$, i.e. maps $A\r K$.
Again, the operation is disjoint union. The example of lax commutative
monoid with cancellation fibered over $S$
considered in \cite{hk} is the groupoid $\mathcal{C}$
of {\em worldsheets} or rigged surfaces. These are 2-dimensional
smooth manifolds with smooth boundary; further, each boundary component
is parametrized by a smooth diffeomorphism with $S^{1}$, and the
surface has a complex structure with respect to which the boundary
parametrization is analytic. Morphisms are biholomorphic diffeomorphisms
commuting with boundary parametrization. 
Addition is disjoint union, and cancellation is gluing of
boundary components. Similarly, again, one can consider the LCMC
$\mathcal{C}_{K}$ of
worldsheets with $K$-labelled boundary components, which is
an LCMC over $S_K$.

\vspace{3mm}
To complete the picture, one needs to consider stacks. We note
that lax algebras over a theory and lax algebras (in our sense) over
a $2$-theory form $2$-categories which have lax limits of strict
diagrams (see Fiore \cite{fioret}). For older references,
which however work in slightly different
contexts (and with different terminology), 
see Borceux \cite{borc} or \cite{fioret}. For the
$2$-category structure, $1$-morphisms are lax morphisms of lax algebras
(functors such that
there is a natural coherence isomorphism for every element of $G$), and
$2$-morphisms are natural isomorphisms which commute with the operations 
given by the theory (or $2$-theory).
Now for any $2$-category $\mathcal{C}$ with lax limits, 
and every Grothendieck topology $\mathcal{B}$,
we can define $\mathcal{B}$-stacks over $\mathcal{C}$: they are simply
contravariant functors $\mathcal{B}\r\mathcal{C}$ which turn Grothendieck
covers into lax limits. Note that such stacks then themselves form a $2$-category
with respect to stack versions of the same $1$-morphisms and $2$-morphisms.

\vspace{3mm}
Now to turn $\mathcal{C}$ into a stack of lax commutative monoids with
cancellation, we must first specify the Grothendieck topology. Note that
there are two choices of the topology, either just (finite-dimensional)
smooth manifolds and open covers (non-chiral setting) or finite-dimensional
complex manifolds and open covers (chiral setting). As remarked in 
Section \ref{s2} above, however, $D$-branes can only be considered in the non-chiral
setting. To define the stack, one must first define the underlying stack 
of lax commutative monoids; the answer is simply the stack of covering
spaces with finitely many sheets. Now one must define smooth or holomorphic
families of worldsheets. We shall only make the definition in the holomorphic
case, the smooth case is analogous.
The most convenient way to make this precise is to consider,
for a worldsheet $X$, the complex $1$-manifold $Y$
obtained by gluing, locally, solid cylinders to the boundary components of $X$.
Then, a holomorphic family of rigged surfaces $X$ over a finite dimensional
complex manifold $B$ is
a holomorphic map 
$$q:Y\r B$$
transverse to every point, such that $dim(Y)=dim(B)+1$ and
$B$ is covered by open sets $U_{i}$ for each of which there are given 
holomorphic regular inclusions
$$s_{i,c}:D\times U_{i}\r Y$$
with
$$q\circ s_{i,c}=Id_{U_{i}}$$
where $c$ runs through some indexing set $C_{i}$. Further, if
$U_{i}\cap U_{j}\neq \emptyset$, we require that there be a bijection
$\iota:C_{i}\r C_{j}$ such that 
$$s_{i,c}|_{D\times(U_{i}\cap U{j})}=s_{j,\iota(c)}|_{D\times(U_{i}\cap 
U_{j})}.$$
Then we let
$$X=Y-(\cform{\bigcup}{i}{} \cform{\bigcup}{c\in C_{i}}{} 
s_{i,c}((D-S^{1})\times
U_{i})).$$
Then the fiber of $X$ over each $b\in B$ is a rigged surface, which vary
holomorphically in $b$, in the sense we want. (Note that the reason
the maps $s_{c}$ cannot be defined globally in $B$ is that it is possible
for a non-trivial loop in $\pi_{1}(B)$ to permute the boundary components
of $X$.) The treatment of $\mathcal{C}_{K}$ is analogous. As a rule,
we shall use the same symbol for the SLCMC's $\mathcal{C}$,
$\mathcal{C}_{K}$ as for the corresponding LCMC's
(their sections over a point).

\vspace{3mm}
We are done with the review of SLCMC's, but we shall still briefly cover
CFT's, as defined in \cite{hk}. Although this definition is subsumed by
Section \ref{s5} above, the reader might still find the more elementary
definition useful while reading the earlier sections.
Let $\mh_{1},...,\mh_{n}$ be complex 
(separable) Hilbert spaces.
Then on $\mh_{1}\otimes...\otimes\mh_{n}$, there is a natural inner
product
$$\langle a_{1}\otimes...\otimes a_{n}, b_{1}\otimes...\otimes b_{n}
\rangle=\langle a_{1},b_{1}\rangle\langle a_{2},b_{2}\rangle...
\langle a_{n},b_{n}\rangle.$$
The Hilbert completion of this inner product space is called
the {\em Hilbert tensor product}
\beg{emht}{\mh_{1}\hat{\otimes}...\hat{\otimes} \mh_{n}.}
Now an element of \rref{emht} is called {\em trace class} if there
exist unit vectors $e_{ij}\in\mh$ where $j=1,..,n$ and $i$ runs through
some countable indexing set $I$ such that
$$x=\cform{\sum}{i\in I}{}\mu_{i}(e_{i1}\otimes...\otimes e_{in})$$
and
$$\cform{\sum}{i\in I}{}|\mu_{i}|<\infty.$$
The vector subspace of \rref{emht} of vectors of trace class will be
denoted by
\beg{embox}{\mh_{1}\boxtimes...\boxtimes\mh_{n}.}
Note that \rref{embox} is not a Hilbert space. We have, however, canonical
maps
$$\boxtimes: (\mh_{1}\boxtimes...\boxtimes\mh_{n})\otimes (\mh_{n+1}
\boxtimes...\boxtimes \mh_{m+n})\r \mh_{1}\boxtimes...\boxtimes\mh_{m+n}$$
and, if $\mh^{*}$ denotes the dual Hilbert space to a complex
Hilbert space $\mh$,
$$tr:\mh\boxtimes\mh^{*}\boxtimes\mh_{1}\boxtimes...\boxtimes\mh_{n}\r
\mh_{1}\boxtimes...\boxtimes\mh_{n}.$$
This allows us to define a particular example of stack of LCMC's based on
$\mh$, which we will call $\underline{\mh}$. The 
underlying stack of lax commutative
monoids ($T$-algebras) is $S$. Now let $B\in \mathcal{B}$. Let $s,t$
be sections of the stack $S$ over $B$, i.e. covering spaces of $B$ with
finitely many sheets. Then we have an infinite-dimensional holomorphic
bundle over $B$
\beg{eboxbundle}{(\mh^{*})^{\boxtimes s}\boxtimes \mh^{\boxtimes t}.}
What we mean by that 
is that there is a well defined sheaf of holomorphic 
sections of \rref{eboxbundle} (note that it suffices to understand the
case when $s$, $t$ are constant covering spaces, which is obvious).
Now a section of $\umh$ over a pair of sections $s$, $t$ of $S$ is
a {\em global} section of \rref{eboxbundle} over $b$; the only
automorphisms of these sections covering $Id_{s}\times Id_{t}$
are identities. The operation $+$, $\check{?}$ are given by the
operations $\boxtimes$, $tr$ (see above).

\vspace{3mm}
We can also define a variation of this LCMC for the case of labels
indexed over a finite set $K$. We need a collection of Hilbert
spaces 
$$\mh_{K}=\{\mh_{k}|k\in K\}.$$
Then we shall define a stack of LCMC's $\umh_{K}$.
The underlying stack of $T$-algebras (commutative monoids) is $S_{K}$.
Let $s,t$ be sections of $S_{K}$ over $B\in\mathcal{B}$.
The place of \rref{eboxbundle} is taken by
\beg{eboxk}{(\mh_{K}^{*})^{\boxtimes s}\boxtimes \mh_{K}^{\boxtimes t}.}
By the sheaf of holomorphic section of \rref{eboxk} when $B$ is a point we
mean that $\boxtimes$-powers of $\mh_{k}$ (or $\mh_{k}^{*}$) for each
label $k\in K$ are taken according to the the number of points of
$\Gamma(t) $ (resp. $\Gamma(s)$); when $s$ and $t$ are constant
covering spaces $B$, the space of sections of \rref{eboxk} is
simply the set of holomorphically varied elements of the spaces
of sections over points of $B$ (which are identified). This is generalized
to the case of general $s$, $t$ in the obvious way (using functoriality
with respect to permutations of coordinates). As above, the only
automorphisms of these sections covering $Id_{s}\times Id_{t}$
are identities.

\vspace{3mm}
\noindent
{\bf Remark:} For technical reasons (different types of convergence), 
the above setup involving Hilbert
spaces and trace class elements is sometimes insufficient (see Example
below and Section \ref{s4} above). Because of that, it is beneficial
to generalize to a context where $\underline{\mathcal{H}}$
simply means any SLCMC over $S$ (resp. $S_{K}$ in the labelled case)
whose spaces of sections are vector spaces, the operation $\check{?}$
is linear, and the operation $+$ is bilinear. We shall further assume
that $\umh$ is a {\em sheaf} in the sense that the only endomorphisms
over the identity in $S$ (resp. $S_{K}$) is the identity.

\vspace{3mm}
\noindent
{\bf Example:}
Consider the free bosonic CFT \rref{efree} discussed in Section \ref{s4}.
As remarked above, the description \ref{efree} is actually 
already not quite right: the vacuum state is to be an eigenstate of
momentum $0$, but there is no non-zero function in $L^{2}(\R,\C)$ with support
in the set $\{0\}$. For the same reason, we also find that the
operator $U_{A_{q}}$ associated with the standard annulus $A_{q}$
is not trace class as defined (since, for example, $1$ is a
limit point of the spectrum of
$U_{A_{q}}$). This is the usual problem in quantum mechanics.
In the present setting, a solution along the lines of the Remark
can be obtained as follows: Let $\mathcal{F}$ be the bosonic Fock space,
i.e.
$$\mf=\widehat{Sym}\langle z^{n},\overline{z}^{n}, n>0\rangle.$$
Then the sections of $\umh$ over $(s,t)$ (over a point) are elements
$$f\in\cform{\prod}{k\in\R^{|s|+|t|}}{}\mf^{\hat{\otimes}|t|}\hat{\otimes}
\mf^{*\hat{\otimes}|s|}$$
(the product is categorical product of vector spaces) which have the
property that for every pair of injections $i:u\r s$, $j:u\r t$
and every map 
$$k:(s-i(u))\coprod (t-j(u))\r \R,$$
we have
$$\cform{\int}{x\in\R^{u}}{}\mu(\phi_{i,j,k}(x))<\infty$$
where $\phi_{i,j,k}(x)=f(y)$, $y\in \R^{s\coprod t}$ is
defined by $y(i(r))=y(j(r))=x(r)$,
$y(r)=k(r)$ for $r\in (s-i(u))\coprod (t-j(u))$ and,
for $z\in\mf^{\hat{\otimes}|t|}\hat{\otimes}\mf^{*\hat{\otimes}|s|}$,
$$\mu(z)=inf(\{\sum|a_{i}| \;|\;z=\sum a_{i}\cform{\otimes}{x\in s\coprod t}{}
e_{x}, ||e_{x}||=1 \;\text{for all $x$}\},$$
and gluing along $i,j$ is defined by
\beg{egglue1}{\check{f}(k)=\cform{\int}{x\in \R^{u}}{}tr(\phi_{i,j,k}(x)).}
The expression \rref{egglue1} is always defined because of the condition
imposed, and the condition is preserved by the gluing operation by Fubini's theorem.

\vspace{3mm}
Now we can define an {\em abstract CFT} based on an SLCMC $\mathcal{D}$ 
with underlying stack of lax commutative monoids (SLCM)
$S$ simply as a $1$-morphism of SLCMC's, over $Id_{S}$,
$$\mathcal{D}\r \underline{\mh}.$$
A similar definition applies if $\mathcal{D}$ has underlying 
SLCM $S_{K}$, with $\underline{\mh}$ replaced by $\underline{\mh}_{K}$.

\vspace{3mm}
However, this notion still is not definitive in the sense that it
does not capture anomaly. In Section \ref{s5} above, we give the
most general definition of modular functor, but it is useful
to review a direct definition from \cite{hk} at least in one
special case, namely $1$-dimensional anomaly.
To this end, we give the definition of {\em $\ct$-central extension}
(or, equivalently, {\em $1$-dimensional modular functor}) on an LCMC
$\mathcal{D}$. This is a strict morphism of stacks of LCMC's 
\beg{epsis}{\psi:\tilde{\mathcal{D}}\r\mathcal{D}}
over $Id$ on the underlying stacks of LCM's
with the following additional structure (for simplicity,
let us just work in the holomorphic (chiral) setting): For each
object $B$ of $\mb$, and each pair of sections $s$, $t$ of $S$ over $B$,
and each section $\alpha$ of $\mathcal{D}$ over $s,t,B$, $B^{\prime}\r B$,
\beg{ecentb}{\psi^{-1}(\alpha|_{B^{\prime}})} 
with varying $B^{\prime}$
is the space of sections of a complex 
holomorphic
line bundle over $B$. Furthermore, functoriality maps supplied by the
structure of stack of LCMC's
on $\tilde{D}$ are linear maps on these holomorphic line
bundles. Regarding the operation $+$, we require that the map
induced by $+$
\beg{ecent+}{\psi^{-1}(\alpha|_{B^{\prime}})\times
\psi^{-1}(\beta|_{B^{\prime}})\r\psi^{-1}((\alpha+\beta)|_{B^{\prime}})}
be a bilinear map, which induces an isomorphism of holomorphic line bundles
\beg{ecent+1}{\psi^{-1}(\alpha|_{B^{\prime}})\otimes_{\mathcal{O}_{B^{\prime}}}
\psi^{-1}(\beta|_{B^{\prime}})\r\psi^{-1}((\alpha+\beta)|_{B^{\prime}})}
($\mathcal{O}_{B}$ is the holomorphic structure sheaf on $B$).

Regarding the operation $\check{?}$, we simply require that if 
$\alpha$ is a section of $\mathcal{D}$ over $s+u,t+u$, $B$
where $u$ is another section of $S$ over $B$, and $\check{\alpha}$ is
the section over $s,t$, $B$ which is obtained by applying
the operation $\check{?}$ to $\alpha$, then the map of
holomorphic line bundles coming
from LCMC structure
\beg{ecentcheck}{\psi^{-1}(\alpha|_{B^{\prime}})\r 
\psi^{-1}(\check{\alpha}|_{B^{\prime}})}
($B^{\prime}\r B$) be an isomorphism of holomorphic line bundles.

By a {\em CFT with $1$-dimensional modular functor over $\mathcal{D}$}
with underlying stack $S$
we shall mean a CFT
\beg{ecenttil}{\phi:\tilde{D}\r\umh}
(where $\tilde{\mathcal{D}}$
is a $\ct$-central extension 
of $\mathcal{D}$
which has the property that $\phi$ is a linear map on the spaces
of sections \rref{ecentb}. Similarly in case $\mathcal{D}$ has underlying
stack $S_{K}$; we simply replace $S$ by $S_{K}$ everywhere throughout
the definition.

\vspace{3mm}
It is appropriate to comment on a weaker kind of morphism of lax algebras
where we do not require that the coherence maps be iso.
By a {\em pseudomorphism} of lax $T$-algebras (and similarly
in the cases of lax $\Xi,T$-algebra
and their stacks) we shall mean a functor
$$f:X\r Y$$
together with morphisms (called cross-morphism, not necessarily iso)
\beg{ecross1}{\gamma(f,...,f)\r f\gamma}
which commute with all the coherences in the lax $T$-algebra sense
(we shall refer to these required commutative diagrams as
cross-diagrams). 

\vspace{3mm}
\noindent
{\bf Remark:}
Although the anomaly of the linear $\sigma$-model considered in Section
\ref{s4} is $1$-dimensional, in Section \ref{s5} we considered higher-dimensional
anomalies. It is therefore appropriate to reconcile 
the above remark concerning generalizing
the SLCMC $\umh$ to cases when the Hibert/trace class model fails due to
non-convergence with our discussion of higher-dimensional modular functors
via $2$-vector spaces. In other words, what is the right generalization of
$H\in Obj(\mathcal{M}^{Hilb})$ in $C(\mathcal{M},H)$?
The main point is that the Hilbert tensor powers $H^{\hat{\otimes}t}
\hat{\otimes}
H^{*\hat{\otimes}s}$ 
should be replaced by a ``vector space indexed
over $\mathcal{M}$'' which depends only on $s,t$ and have appropriate designated
``trace maps''. 

The category $\mathcal{M}^{Vect}$ of {\em vector spaces indexed over $\mathcal{M}$}
is defined as
$$\mathcal{M}\otimes_{\C_{2}}\C_{2}^{Vect}$$ 
where $\C_{2}^{Vect}$ is the lax commutative semiring of $\C$-vector spaces
(not necessarily finitely dimensional). 

Now $H\in Obj(\mathcal{M}^{Vect})$, we may consider a pseudomorphism
of SLCMC's over $S$ (see above) 
$$h:S^{2}\r C(\mathcal{M}^{Vect}).$$
Here by $C(\mathcal{M}^{Vect})$ we mean the analogous construction as 
$C(\mathcal{M})$, but with the duals taken over $\C^{Vect}$, so
$(\mathcal{M}^{Vect})^{*}=_{def}(\mathcal{M}^{*})^{Vect}$.
Then, an SLCMC $C(\mathcal{M}, h)$  is defined as follows. Sections over
a point over $\sigma\in Obj(S)^{2}$ consist of a section $M$ of
$C(\mathcal{M})$ over $\sigma$ and a $2$-morphism
$$M\r h(\sigma).$$
Stacking, and the necessary verifications, are completed in the usual way.


\vspace{15mm}
\begin{thebibliography}{99}

\bibitem{ando} M. Ando: Power operations in elliptic cohomology and representations
of loop groups, {\em Trans. AMS} 352 (2000) 5619-5666

\bibitem{rog} N.A.Baas, B.I.Dundas, J.Rognes: Two-vector bundles and
forms of elliptic cohomology, to appear


\bibitem{borc} F.Borceux: {\em Handbook of categorical algebra 1-2},
Encyclopedia of Mathematics and its Applications 50-52, Cambridge
University Press

\bibitem{borch} R.E.Borcherds: Monstrous moonshine and monstrous Lie
superalgebra, {\em Invent. Math.} 109 (1992) 405-444

\bibitem{dh}deHoeker: String theory, in {\em Quantum fields and strings:
a course for mathematicians}, vol. 2, AMS and IAS 1999, 807-1012

\bibitem{df} P.Deligne, D.Freed: Notes on Supersymmetry
(following J. Bernstein) in: {\em Quantum fields and strings,
a course for mathematicians}, vol. 1, AMS, 1999, 41-98

\bibitem{diac} D.E.Diaconescu: Enhanced $D$-brane categories from
string field theory, {\em JHEP} 0106 (2001) 16

\bibitem{doug} M.R.Douglas: $D$-branes, categories and $N=1$ SUSY,
{\em J.Math.Phys.} 42 (2001) 2818-2843

\bibitem{fioret} T.Fiore: Lax limits, lax adjoints and lax algebras,
preprint, 2003

\bibitem{f} I.Frenkel: Vertex algebras and algebraic curves, Seminaire Bourbaki
1999-2000,
{\em Asterisque} 276 (2002) 299-339

\bibitem{flm} I.Frenkel, J.Lepowsky, A.Meurman: {\em
Vertex operator algebras and the monster}, Pure and applied Mathematics,
vol. 134, Academic Press, 1999

\bibitem{gsw} M.B.Green, J.H.Schwartz, E.Witten: {\em Superstring theory},
vol. 1,2, Cambridge University Press, 1988

\bibitem{hor} P.Horava: Equivariant Topological Sigma Models, Nucl. Phys. B 418 (1994)
571-602, hep-th/9309124

\bibitem{hk} P.Hu, I.Kriz: Conformal field theory and elliptic cohomology,
to appear in {\em Advances in Mathematics}

\bibitem{hkv} P.Hu, I.Kriz, A.A.Voronov: On Kotsevich's Hochschild cohomology
conjecture, Preprint 2001

\bibitem{spin} I.Kriz: On spin and modularity in conformal field theory,
{\em Ann. Sci. de ENS} 36 (2003) 57-112

\bibitem{law} W.F.Lawvere: Functorial semantics of algebraic theories,
{\em Proc. Nat. Acad. Sci. U.S.A.} 50 1963 869--87

\bibitem{laz} C.I.Lazaroiu: On the structure of open-closed topological
field theory in two-dimensions, {\em Nucl. Phys. B} 603 (2001) 497-530

\bibitem{laz2} C.I.Lazaroiu: Generalized complexes and string field
theory, {\em JHEP} 0106 (2001) 52

\bibitem{laz3} C.I.Lazaroiu: Unitarity, $D$-brane dynamics an $D$-brane
categories, {\em JHEP} 0112 (2001) 31

\bibitem{lew} D.Lewellen: Sewing constraints for conformal field theories
on surfaces with boundaries, Nucl. Phys. B 372 (1992) 654

\bibitem{ms} G.Moore: Some Comments on Branes, $G$-flux, and $K$-theory, 
{\em Int. J.Mod.Phys.} A16 (2001) 936-944

\bibitem{ms1} G.Moore, N.Seiberg: Classical and Quantum Conformal Field Theory,
{\em Comm. Math. Phys.} 123, (1989) 177-254 

\bibitem{ms2} G.Moore, N.Seiberg: Taming the conformal ZOO, {\em Physics Letters B}
220 (1989) 422-430

\bibitem{ms3} G.Moore, N.Seiberg: Lectures on RCFT, {\em Physics, Geometry and Topology},
H.C.Lee, ed., (1990) 263-361

\bibitem{pol} J.Polchinski: {\em String theory}, Vol. 1,2, Cambridge Univ. Press, 1999

\bibitem{ps} A.Pressley, G.Segal: {\em Loop groups}, Oxford University Press, 1986

\bibitem{s} G.Segal: Elliptic cohomology, Seminaire Bourbaki 1987/88,
{\em Asterisque} 161-162 (1988) Exp. No, 695, (1989) 4, 187-201 

\bibitem{scft} G. Segal: The definition of conformal field theory, Preprint, the 1980's

\bibitem{seg} G.Segal: ITP lectures, http://doug-pc.itp.ucsb.edu/online/geom99/

\bibitem{sc} G.Segal: Categories and cohomology theories, {\em Topology}
13 (1974), 293-312

\bibitem{st} S.Stolz, P.Teichner: What is an elliptic object?, preprint, 2003

\bibitem{thomason} R.W.Thomason: Beware the phony multiplication on Quillen's
$\mathcal{A}^{-1}\mathcal{A}$, {\em Proc. AMS} 80 (1980) 4, 569-573

\bibitem{verlinde} E.Verlinde: Fusion rules and modular transformations
in $2D$ conformal field theory, {\em Nucl. Phys. B} 300 (1988) 360-376

\bibitem{w} E.Witten: Overview of $K$-theory applied to strings, 
{\em Int.J.Mod.Phys.} A 16 (2001) 693-706


\end{thebibliography}
\end{document}